\documentclass[journal,comsoc]{IEEEtran}
\usepackage[T1]{fontenc}
\usepackage{cite}
\usepackage{amsmath,amssymb,amsfonts}
\usepackage{amsthm}
\usepackage{algorithmic}
\usepackage{algorithm}
\usepackage{graphicx}
\usepackage{textcomp}
\usepackage{xcolor}
\usepackage{graphicx}
\usepackage{bm,tikz,pgfplots,stmaryrd,wrapfig}
\usepackage{booktabs} 
\usepackage{multirow}
\usetikzlibrary{positioning,arrows,arrows.meta,shapes.geometric,calc,decorations.pathreplacing}
\tikzstyle{block} = [draw, rectangle, minimum height=1em, minimum width=2em,text centered]
\tikzstyle{lblock} = [draw, rectangle, minimum height=2em, minimum width=5cm,text centered]
\tikzstyle{summer} = [draw, circle, text centered]
\tikzstyle{attenuatorR} = [draw, regular polygon, regular polygon sides=3,shape border rotate=30, text centered]
\tikzstyle{antenna} = [draw, regular polygon, regular polygon sides=3,shape border rotate=180, text centered]

\def\BibTeX{{\rm B\kern-.05em{\sc i\kern-.025em b}\kern-.08em
    T\kern-.1667em\lower.7ex\hbox{E}\kern-.125emX}}

 \pgfplotsset{compat=1.17}

\ifCLASSINFOpdf
\else
\fi
\usepackage{amsmath}
\usepackage[cmintegrals]{newtxmath}
\begin{document}
%
\title{Robust Statistical Beamforming with Multi-Cluster Tracking for Time-Varying Massive MIMO (Extended Version)}
%
%
%

\author{Anil~Kurt
        and~Gokhan~M.~Guvensen
\thanks{The authors are with the Department of Electrical and Electronics Engineering, Middle East Technical University, Ankara, Turkey (e-mail:
anilkurt@metu.edu.tr; guvensen@metu.edu.tr).}
}

%
%

\markboth{March~2023}%
{Shell \MakeLowercase{\textit{et al.}}: Bare Demo of IEEEtran.cls for IEEE Communications Society Journals}
%



\maketitle

\begin{abstract}
In this paper, a joint design of instantaneous channel estimation, beam tracking, and adaptive beamformer construction for a massive multiple-input multiple-output (MIMO) system is proposed. This design focuses on efficiency in terms of performance and computational complexity under the adverse effects of time variation and mobility of sources, the presence of multiuser and multipath components, or simply multi-clusters, and the near-far effect. The design is also suitable for hybrid beamforming and frequency-selective channels. In the proposed system, channel parameters are estimated in time-domain duplex (TDD) uplink mode using a per-cluster approach rather than a joint approach, which significantly reduces the complexity. Per-cluster estimation is possible thanks to the proposed interference-aware statistical beamforming method, namely reduced dimensional Generalized Eigenbeamformer (RD-GEB), which undertakes the computational load of interference mitigation and enables a simpler design for the remaining stages. In addition, the overall design is based on the separation of channel parameters as \emph{fast-time} and \emph{slow-time}, leaving only the instantaneous channel estimation and channel matched filtering as fast-time operations, which are handled inside cluster-specific reduced dimensional subspaces. Beam tracking and beamformer construction are held in slow-time rarely, which reduces the time-averaged complexity. Furthermore, beam tracking is performed by leveraging a batch of instantaneous channel estimates, which removes the need for an additional training process. The proposed low-complexity design is shown to outperform the conventional methods.
\end{abstract}

\begin{IEEEkeywords}
Statistical beamforming, time variation, mobility, multiuser, interference mitigation, beam tracking
\end{IEEEkeywords}

%
\IEEEpeerreviewmaketitle

\section{Introduction}
%
%
%
%
\IEEEPARstart{M}{assive} multiple-input multiple-output (MIMO) is a widely studied technology that is the fundamental basis for new-generation communication systems \cite{Larsson14}. Using large antenna arrays, massive MIMO increases the data rate, brings angular selectivity and angular focusing of the radiated power, and compensates for huge path loss in millimeter-wave communications \cite{Lu14}. It also enables spatial multiplexing of multiple users, removing the need for temporal or spectral division \cite{Sun15,Jia19}. Studies show that it increases the performance of full-duplex communications \cite{Ngo14} and radar applications \cite{Fortunati20}. In addition, it is the basis for various new applications such as dual-functional radar-communication (DFRC) \cite{Liu20}, reconfigurable intelligent surfaces (RIS) \cite{Huang20}, and cell-free massive MIMO \cite{Ngo16}, and a myriad of 6G technologies such as high-speed train (HST) and vehicle-to-vehicle (V2V) communications \cite{Wang20}. 

As technology evolves, the number of actively communicating devices is expected to grow significantly \cite{Wang20}. In a time-varying scenario where multiple users are mobile, besides serving, channel estimation is a significant problem that results in explore/exploit dilemma and a huge computational load. 

The hybrid beamformer (HBF) structure also complicates the estimation procedure. HBF is a widely offered structure for large antenna arrays instead of fully digital or fully analog beamformers for its efficiency regarding performance and physical viability \cite{Heath16}. It includes both analog beamformer (ABF) and digital beamformer (DBF) stages, with a reduced number of radio-frequency chains (RFCs). Since the ABF always needs to be directed to some angle(s), and angle of arrival (AoA) estimation is held in the digital stage after the ABF, AoA estimation is limited by the prior ABF beam coverage. In addition, user mobility might result in beam loss or unintended suppression due to ABF. These factors obligate either a scanning operation or a beam tracking procedure. 
%

Computational complexity is a significant factor affecting the feasibility of a system design. In time-varying systems, varying channel parameters should be estimated frequently and systems using these parameters in their constructions should be updated accordingly. These estimation and update operations should be repeated within the interval in which the related parameters can be assumed to remain almost the same, namely the coherence time. On the other hand, the main task that brings the most complexity is interference mitigation, which requires inversion of matrices in the majority of techniques. Interference might be caused by the presence of multiple users and multipath components, or simply multiple signal clusters in a MIMO system.

For time-varying channels, the rate of variation imposed on channel parameters might be different \cite{AngleGainCT}. Among the channel parameters, slowly varying and rapidly varying ones are often called slow-time (ST) and fast-time (FT) parameters, respectively. In the massive MIMO case, AoA and angular spread (AS) could be classified as ST parameters, while gains of individual micro-components inside a signal cluster could be FT parameters. Based on this classification, instantaneous channel estimation should be repeated after shorter intervals (in fast-time), while beam tracking can be repeated more rarely (in slow-time). Upon the estimates from these procedures, beamformer weights can be updated either in fast-time or slow-time depending on the method that is used, namely instantaneous or statistical beamforming.

\subsection{Literature Review} \label{sec_literature}
Angular estimation for massive MIMO systems is a widely studied topic. While exhaustive or hierarchical search is offered for initial estimation \cite{Ma22, Shaham19, Palacios17, Alkhateeb15, Liu17}, lower-complexity beam tracking operations are offered after the initial phase to adaptively modify the estimate under variations through time. 

In the vast majority of the studies, beam tracking is held after ABF consisting of phase shifters, which create discrete Fourier transform (DFT) beams, to probe angles in space. In \cite{Gao17, Zhang19, Zhang21, Hussain22, Ma22}, beam tracking means the selection of the optimum set of AoAs to be probed under time or physical (number of RFCs) limitations, after which the most powerful DFT beams are selected. The use of two perturbed beams around the main AoA is also common \cite{Li21, Liu21, Zhu18}, from which the final AoA estimate is calculated accordingly. In the aforementioned studies, beam tracking is a phase where the beamformer is varied for a search. On the contrary, AoA variation is deduced from the variations on the current ABF output in \cite{Shaham19, Va16, Jayaprakasam17} with the help of extended Kalman filter (EKF), and in \cite{Palacios17} via maximum a posteriori (MAP) estimator. \cite{Zhu22} and \cite{Palacios17} employ fully digital beamformers, and \cite{Kurt20} assumes erroneous AoA estimates and tracks the beams using recursive filtering.

Beam tracking in multiuser channels is investigated in only a few studies. Further, the near-far effect, to the authors' knowledge, is not studied in the literature in the beam tracking context. That is the reason why the ABF output, whose sidelobe suppression is limited, is generally chosen for beam tracking. Against multiuser interference, \cite{Zhao17} proposes a user scheduling algorithm, \cite{Zhu22} employs orthogonal pilots, and \cite{Gao17} and \cite{Zhu18} offer time or code domain multiplexing. In \cite{Zhang21}, multiuser estimation is performed either jointly by extending the MAP estimator, or sequentially by successive interference cancellation. \cite{Li21} also applies joint estimation after ABF with the help of a maximum likelihood (ML) estimator. 

Contrary to the algorithms that operate after ABF, \cite{Ma22} proposes measurement after a digital zero-forcing (ZF) beamformer after ABF, to suppress the multiuser interference further. However, since ZF is constructed with the channel estimates, it is, as we will define later, an \emph{instantaneous beamformer}, which obligates frequent training phases. Indeed, beam tracking is generally mixed with the estimation of the rapidly changing channel gain coefficients in the literature. In \cite{Shaham19, Va16, Jayaprakasam17, Li21, Zhang21}, the proposed algorithms track the gain coefficient together with the AoA. On the contrary, \cite{AngleGainCT} defines gain coherence time and angle coherence time such that the angle coherence time is much longer. Furthermore, \cite{Liu21} studies the optimum period for beam tracking.

%
Training sequence transmission is needed for beam tracking in most studies. Although uplink training is proposed generally, \cite{Zhang19, Zhang21, Hussain22, Zhu18} propose downlink training at the user equipment (UE), which requires feedback to the base station (BS). In \cite{Ma22}, different DFT beams are tried inside the data mode, accepting data rate losses during the process.

Besides conventional communications, beam tracking is an important topic also for evolving technologies. It is studied for RIS \cite{Hashida22}, unmanned air vehicles (UAVs) \cite{Chang22}, automated vehicles \cite{Zhang22}, and terahertz communications \cite{Ning22}.

\subsection{Contributions}
In this paper, a novel per-cluster estimation scheme is proposed for massive MIMO systems, which requires an inclusive consideration of the whole beamformer structure, beamforming approach, physical constraints, computational complexity, and time scheduling. 

A statistical beamformer, namely generalized eigenbeamformer (GEB) from our previous work \cite{Kurt19}, is adapted to HBFs, which have a bank of analog phase shifters for ABF, to be used in the digital stage with the name reduced dimensional (RD) GEB (RD-GEB). Statistical beamformers do not depend on instantaneous channel coefficients, and they provide wider selection and null areas in the angular domain, based on the channel model that includes AS. Therefore, the validity of the beamformer lasts longer, which enables the definition of a separate coherent processing interval (CPI) for beam tracking, namely the slow-time CPI (ST-CPI), which is longer than the CPI for the instantaneous channel coefficients, fast-time CPI (FT-CPI). Moreover, it mitigates the multipath and multiuser interference better than widely proposed DFT beams, and creates interference-free subspaces for all clusters even under strong near-far effects. Consequently, multiuser channel estimation and beam tracking are performed via angular division in these subspaces in time-domain duplex (TDD) uplink mode, namely \emph{per-cluster}, without spending resources via user scheduling algorithms, orthogonal pilot transmissions, or time division/synchronization between users. Further, the computational complexities of these estimation operations are very low because interference mitigation is handled previously by the statistical beamformer, and the dimension is further reduced. Also, beamformer construction and beam tracking are performed rarely in slow-time, but the instantaneous effective channel (IEC) estimator operates frequently with RD inputs in fast-time. Since the dominant complexity of the interference mitigation is undertaken by the ST beamformer, and the FT estimator is of low complexity, the time-averaged computational complexity also reduces. The indifference of the statistical beamformer to instantaneous variations of the channel is handled by a simple intra-cluster spatial channel matched filter (ICS-CMF), which is matched to the IEC.

The second major contribution is the design of novel beam tracking methods, namely beam-aware maximum likelihood estimator (BA-ML) and statistical extended Kalman filter (SEKF). Besides the fact that they both operate per-cluster after the statistical beamformer, both leverage the IEC estimates collected throughout an ST-CPI. Therefore, there is no need for an additional period of training. The difficulty of utilizing multiple channel estimates due to uncorrelated channel gains is handled meticulously in the design. While the BA-ML method is a variant of the nonlinear least squares method \cite{stoica}, SEKF is an EKF application whose observation is the second-order statistics, namely the effective channel covariance matrix (CCM). To sum up, the main contributions of this paper are listed below.
\begin{itemize}
\item The design of a novel per-cluster estimation approach with statistical beamforming, whose advantages are
\begin{itemize}
\item Suitability to
\begin{itemize}
\item hybrid beamformers (HBFs),
\item multiuser channels under notable near-far effect,
\item time-varying channels and mobility,
\item frequency-selective channels.
\end{itemize}
\item High multiuser estimation performance thanks to the reduced dimensional interference-free subspaces, which also enables the usage of simpler estimators designed for single-user cases,
\item Per-cluster estimation which removes the need for synchronization between the users, 
\item Low overall computational complexity thanks to the careful distribution of complexity load between FT and ST blocks, (especially the highly complex task of multiuser interference mitigation handled by ST statistical beamformer)
\item Reduced delay spread for IEC estimation compared with joint estimation.
\end{itemize}
\item The design of novel beam tracking methods, namely BA-ML and SEKF, which
\begin{itemize}
    \item leverage the collection of IEC estimates,
    \item remove the need for an additional training period.
\end{itemize}
\end{itemize}

In addition, we adapt techniques from our prior work \cite{Kurt19} to support the per-cluster estimation scheme for the remaining parts of the system, which are RD-GEB for beamformer construction, and beam-aware least squares (LS, BA-LS) for IEC estimation. Also, Orthogonal Matching Pursuit (OMP) method is adapted as a beam tracker to represent a joint estimation method in order to compare with per-cluster BA-ML and SEKF methods.

In the remainder of the paper, subscripts and superscripts generally indicate the time and cluster affiliations, respectively. Also, $\bar{x}$ and $\tilde{x}$ indicate the transform of an arbitrary variable $x$ after ABF and DBF, respectively. Also, $(\bm{X})_{a,b}$ indicates the entry of $\bm{X}$ in $a$\textsuperscript{th} row and $b$\textsuperscript{th} column. Finally, $\bm{x}^{H}$ and $\bm{x}^{T}$ indicate the Hermitian and transpose operations, respectively.

\section{System Model}
This study considers a massive MIMO system where $U$ single-antenna mobile UEs are simultaneously communicating with a BS with a uniform linear array (ULA) of $N$ antenna elements in TDD uplink mode using single-carrier modulation. Signals arrive at BS as $M$ angularly resolved signal clusters, where $M~\ge~U$. Parameters of MIMO channels are grouped into two, namely FT and ST parameters. The BS estimates these parameters in separate repetition intervals, namely FT-CPI and ST-CPI. The estimated channel parameters can also be used in TDD downlink mode leveraging the channel reciprocity. The most important variables that will be defined in the subsequent sections are listed in Table \ref{tab_nomenclature}.

\begin{table}[tb]
\caption{\textsc{Nomenclature}}
	\centering
	\resizebox{0.48\textwidth}{!}{%
 \bgroup
\def\arraystretch{1.5}
	\begin{tabular}{|r|l|}
	    \hline
		Variable & Description 
        \\ \hline \hline
		$n$, $p$, $k$, $m$, $u$ & Index for symbol, FT-CPI, ST-CPI, cluster, user
        \\ \hline
        $N$, $R$, $D_m$ & \# Antennas, RFCs (ABF outputs), DBF outputs ($\triangleq$ *)
        \\ \hline
        $P$, $M$, $U$ & \# FT-CPIs in an ST-CPI, clusters, users
        \\ \hline
        $N_{\text{F}}$, $N_{\text{S}}$ & \# Symbols in (training, data) mode in an FT-CPI,
        \\ \hline
		$\bm{S}$,  $\bm{W}^{(m)}$,  $\bm{T}^{(m)}$ & ABF, DBF, Total BF matrices
        \\ \hline
		$\bm{y}_{n,p}$,  $\bm{r}_{n,p}$,  $\bm{z}_{n,p}^{(m)}$ & Received Signal (at *)
        \\ \hline
		$\bm{h}_p^{(m)}$,  $\bar{\bm{h}}_p^{(m)}$,  $\tilde{\bm{h}}_p^{(m,m)}$ & Channel (at *)
        \\ \hline
		$s_{n,p}^{(m)}$,  $b_{n,p}^{(u)}$ & Symbols from clusters and users
        \\ \hline
		$E^{(m)}$, $N_0$ & Cluster power, noise variance
        \\ \hline
            $\alpha_{l,p}^{(m)}$,  $\bm{a}(\theta)$ & Channel complex gain, ULA steering vector
        \\ \hline
		$\bm{R}_p^{(m)}$,  $\bar{\bm{R}}_p^{(m)}$,  $\tilde{\bm{R}}_p^{(m,m)}$ & Channel covariance matrix (at *)
        \\ \hline
		$\bm{\Psi}_p$,  $\bar{\bm{\Psi}}_p$,  $\tilde{\bm{\Psi}}_p^{(m)}$ & Total covariance matrix (at *)
        \\ \hline
		$\bm{x}_p^{(m)}$,  $\bm{\nu}_p^{(m)}$,  $\bm{A}$ & Mobility state vector, innovation, transition matrix
        \\ \hline
		$\theta_p^{(m)}$,  $\Delta_p^{(m)}$ & Mean AoA, AS
        \\ \hline
		$\hat{\tilde{\bm{h}}}_p^{(m)}$,  $\bm{\xi}_p^{(m)}$ & IEC estimate, estimation error
        \\ \hline 
             $\bm{R}_k^{(\text{ST},m)}$,  $\bm{\Psi}_k^{(\text{ST})}$, $\bar{\bm{R}}_k^{(\text{ST},m)}$,  & Some ST model variables ( $(\cdot)_k^{(\text{ST})}$ ) corresponding
        \\ $\bar{\bm{\Psi}}_k^{(\text{ST})}$, $\bm{x}_k^{(\text{ST},m)}$, $\theta_k^{(\text{ST},m)}$ & to the previously defined variables ( $(\cdot)_p$ )
  \\ \hline 
	\end{tabular}
 \egroup
	}
	\label{tab_nomenclature}			
\end{table}

\subsection{Signal Model for Single Carrier Uplink Transmission}
In TDD uplink mode, the received signal vector $\bm{y}_{n,p} \in \mathbb{C}^{N\times 1}$ in $n$\textsuperscript{th} discrete time in $p$\textsuperscript{th} FT-CPI is
\begin{equation} \label{eq_rec_signal}
    \bm{y}_{n,p} = \sum_{m=1}^{M} \sqrt{E^{(m)}} \bm{h}_{p}^{(m)} s_{n,p}^{(m)} + \bm{\eta}_{n,p}
\end{equation}
where $s_{n,p}^{(m)}$ are zero-mean unit-variance symbols, $E^{(m)}$ are cluster powers, and $\bm{h}_{p}^{(m)} \in \mathbb{C}^{N \times 1}$ is the channel vector for $m$\textsuperscript{th} cluster. Also, $\bm{\eta}_{n,p} \sim \mathcal{CN} (\bm{0},N_0 \bm{I}_N)$ is the AWGN vector.

Signal clusters might belong to $U$ different users with $U=M$, or some clusters might be multipath components from a less number of users with $U < M$. This fact can be shown as
\begin{equation} \label{eq_cluster2user}
    s_{n,p}^{(m)}=b_{n-l_m,p}^{(\mathcal{U}(m))},
\end{equation}
where $b_{n,p}^{(u)}$ is the transmitted symbol from $u$\textsuperscript{th} user, and $l_m$ indicates the delay for $m$\textsuperscript{th} cluster. $\mathcal{U}(m)$ is the user-cluster association function from the set of clusters $\{1,\dots,M\}$ to the set of users $\{1,\dots,U\}$. It is simply an identity function $\mathcal{U}(m)=m$ for the case $U=M$.

\begin{figure}[tb]
\centering
\resizebox{0.48\textwidth}{!}{%
\begin{tikzpicture}[>=triangle 45]

\def\xsc{1}  
\def\ysc{0.3*\xsc} 

\def\yA{0}
\def\yscA{\ysc}
\def\yB{-6.5*\ysc}
\def\yscB{\ysc}
\def\yC{-15*\ysc}
\def\yscC{\ysc}

\def\xA{0}
\def\xscA{\xsc}

\def\xscB{1.5*\xsc}
\def\xB{0}

\def\xscC{1.1*\xsc}
\def\xC{0}

\def\xAz{\xA - 1*\xscA}
\def\xAy{\xAz - 0.2*\xscA}
\def\xAx{\xAy - 0.2*\xscA}
\def\xAw{\xAx - 1*\xscA}
\def\xAu{\xAw - 1*\xscA}
\def\xAt{\xAu - 0.2*\xscA}
\def\xAs{\xAt - 0.2*\xscA}
\def\xAr{\xAs - 1*\xscA}
\def\xAp{\xAr - 0.5*\xscA}

\def\xAa{\xA + 1*\xscA}
\def\xAb{\xAa + 0.2*\xscA}
\def\xAc{\xAb + 0.2*\xscA}
\def\xAd{\xAc + 1*\xscA}
\def\xAe{\xAd + 1*\xscA}
\def\xAf{\xAe + 0.2*\xscA}
\def\xAg{\xAf + 0.2*\xscA}
\def\xAh{\xAg + 1*\xscA}
\def\xAi{\xAh + 0.5*\xscA}

\def\yAz{\yA + 1*\yscA}
\def\yAa{\yA - 1*\yscA}
\def\yAba{\yAa - 0.5*\yscA}
\def\yAb{\yAba - 0.5*\yscA}
\def\yAbc{\yAb - 0.5*\yscA}
\def\yAc{\yAbc - 0.5*\yscA}

\node (STCPI) [anchor=south,align=center] at (\xA,\yAz) {FT CPIs ($p=1 \rightarrow \infty$) \& ST CPIs ($k=\lceil p/P \rceil : 1 \rightarrow \infty$)};

\draw (\xAr,\yAa) -- (\xAh,\yAa);
\draw (\xAr,\yAc) -- (\xAh,\yAc);

\draw (\xAs,\yAa) -- (\xAs,\yAc);
\draw (\xAs,\yAba) -- (\xAu,\yAba);
\draw (\xAs,\yAb) -- (\xAu,\yAb);
\draw (\xAs,\yAbc) -- (\xAu,\yAbc);

\draw (\xAu,\yAa) -- (\xAu,\yAc);
\draw (\xAx,\yAa) -- (\xAx,\yAc);
\draw (\xAx,\yAba) -- (\xAz,\yAba);
\draw (\xAx,\yAb) -- (\xAz,\yAb);
\draw (\xAx,\yAbc) -- (\xAz,\yAbc);

\draw (\xAz,\yAa) -- (\xAz,\yAc);
\draw (\xAa,\yAa) -- (\xAa,\yAc);
\draw (\xAa,\yAba) -- (\xAc,\yAba);
\draw (\xAa,\yAb) -- (\xAc,\yAb);
\draw (\xAa,\yAbc) -- (\xAc,\yAbc);

\draw (\xAc,\yAa) -- (\xAc,\yAc);
\draw (\xAe,\yAa) -- (\xAe,\yAc);
\draw (\xAe,\yAba) -- (\xAg,\yAba);
\draw (\xAe,\yAb) -- (\xAg,\yAb);
\draw (\xAe,\yAbc) -- (\xAg,\yAbc);

\draw (\xAg,\yAa) -- (\xAg,\yAc);

\draw[dashed] (\xAp,\yAa) -- (\xAr,\yAa);
\draw[dashed] (\xAp,\yAc) -- (\xAr,\yAc);
\draw[dashed] (\xAh,\yAa) -- (\xAi,\yAa);
\draw[dashed] (\xAh,\yAc) -- (\xAi,\yAc);
\node () [] at (\xAr,\yAb) {$p=1,2,\dots$};
\node () [] at (\xAh,\yAb) {$p \rightarrow \infty$};

\draw[<->] (\xAu,\yA) -- (\xAz,\yA);
\draw[<->] (\xAz,\yA) -- (\xAc,\yA);
\draw[<->] (\xAc,\yA) -- (\xAg,\yA);

\node (stz) [fill=white] at (\xAw,\yA) {$k-1$};
\node (sta) [fill=white] at (\xA,\yA) {$k$};
\node (stb) [fill=white] at (\xAd,\yA) {$k+1$};


\def\xBzz{\xB - 0.2*\xscB}
\def\xBz{\xB - 0.4*\xscB}
\def\xBy{\xBz - 1*\xscB}
\def\xBx{\xBy - 1*\xscB}
\def\xBw{\xBx - 0.5*\xscB}
\def\xBu{\xBw - 0.5*\xscB}

\def\xBaa{\xB + 0.2*\xscB}
\def\xBa{\xB + 0.4*\xscB}
\def\xBb{\xBa + 1*\xscB}
\def\xBc{\xBb + 1*\xscB}
\def\xBd{\xBc + 0.5*\xscB}
\def\xBe{\xBd + 0.5*\xscB}

\def\yBa{\yB - 1*\yscB}
\def\yBba{\yBa - 0.5*\yscB}
\def\yBb{\yBba - 0.5*\yscB}
\def\yBbc{\yBb - 0.5*\yscB}
\def\yBc{\yBbc - 0.5*\yscB}

\node (FTCPIs) [anchor=south, align=center] at (\xA,\yBa) {The $k$\textsuperscript{th} ST-CPI (Slow-Time CPI)};

\draw (\xBu,\yBa) -- (\xBzz,\yBa);
\draw[dashed] (\xBzz,\yBa) -- (\xBaa,\yBa);
\draw (\xBaa,\yBa) -- (\xBe,\yBa);
\draw (\xBu,\yBc) -- (\xBzz,\yBc);
\draw[dashed] (\xBzz,\yBc) -- (\xBaa,\yBc);
\draw (\xBaa,\yBc) -- (\xBe,\yBc);

\draw (\xBu,\yBa) -- (\xBu,\yBc);
\draw (\xBu,\yBba) -- (\xBx,\yBba);
\draw (\xBu,\yBb) -- (\xBx,\yBb);
\draw (\xBu,\yBbc) -- (\xBx,\yBbc);
\draw (\xBx,\yBa) -- (\xBx,\yBc);
\draw (\xBz,\yBa) -- (\xBz,\yBc);
\draw (\xBa,\yBa) -- (\xBa,\yBc);
\draw (\xBc,\yBa) -- (\xBc,\yBc);
\draw (\xBe,\yBa) -- (\xBe,\yBc);
\draw (\xBc,\yBba) -- (\xBe,\yBba);
\draw (\xBc,\yBb) -- (\xBe,\yBb);
\draw (\xBc,\yBbc) -- (\xBe,\yBbc);

\node () [] at (\xBy,\yBb) {$p=(k-1)P+1$};
\node () [] at (\xB,\yBb) {$\cdots$};
\node () [] at (\xBb,\yBb) {$p=kP$};
\node () [align=center,anchor=north] at (\xBw,\yBc) {$k-1$\textsuperscript{th} \\ ST beam tracking \\ \& BF update};
\node () [align=center,anchor=north] at (\xBd,\yBc) {$k$\textsuperscript{th} \\ ST beam tracking \\ \& BF update};

\def\xCz{\xC - 2*\xscC}
\def\xCy{\xCz - 1*\xscC}
\def\xCx{\xCy - 1*\xscC}
\def\xCa{\xC + 1*\xscC}
\def\xCb{\xCa + 3*\xscC}
\def\yCa{\yC - 1*\yscC}
\def\yCba{\yCa - 0.5*\yscC}
\def\yCb{\yCba - 0.5*\yscC}
\def\yCbc{\yCb - 0.5*\yscC}
\def\yCc{\yCbc - 0.5*\yscC}
\def\yCd{\yCc - 1*\yscC}
\def\yCe{\yCd - 1*\yscC}

\node () [anchor=south] at (\xC,\yCa) {An FT-CPI (Fast-Time CPI)};
\draw (\xCx,\yCa) -- (\xCb,\yCa);
\draw (\xCx,\yCd) -- (\xCb,\yCd);

\draw (\xCx,\yCa) -- (\xCx,\yCd);
\draw (\xCz,\yCa) -- (\xCz,\yCd);
\draw (\xCb,\yCa) -- (\xCb,\yCd);
\node () [align=center] at (\xCy,\yCbc) {FT Training\\ ($N_{\text{F}}$ symbols)};
\node () [align=center] at (\xCa,\yCbc) {Data Mode: $n=1,\dots,N_{\text{S}}$\\ ($N_{\text{S}}$ symbols)};
\draw[<->] (\xCx,\yCe) -- (\xCb,\yCe);
\node () [fill=white] at (\xC,\yCe) {$T_{\text{F}}$ seconds};
\draw[dotted] (\xAx,\yAc) -- (\xBu,\yBa);
\draw[dotted] (\xAc,\yAc) -- (\xBe,\yBa);
\draw[dotted] (\xBx,\yBc) -- (\xCx,\yCa);
\draw[dotted] (\xBz,\yBc) -- (\xCb,\yCa);
\end{tikzpicture}
}
\caption{Signaling schemes and coherent processing intervals.}
\label{fig_cpi}
\end{figure}
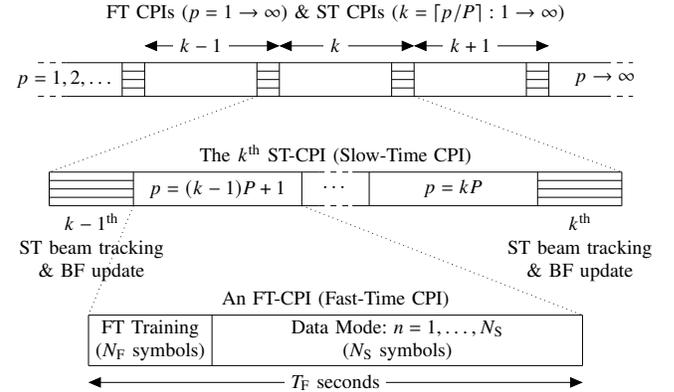

\subsection{Multi-Cluster Massive MIMO Channel Model}
In the $p$\textsuperscript{th} FT-CPI, the channel model for the $m$\textsuperscript{th} cluster is
\begin{equation} \label{eq_channel}
    \bm{h}_{p}^{(m)} = \frac{1}{\sqrt{L}} \sum_{l=1}^L \alpha_{l,p}^{(m)} \bm{a}(\tilde{\theta}_{l,p}^{(m)})
\end{equation}
where $\bm{a}(\theta)\triangleq \frac{1}{\sqrt{N}}[1 \, e^{j \pi \sin{\theta}} \cdots e^{j (N-1) \pi \sin{\theta}}]$, and $L$ is the number of rays. $\alpha_{l,p}^{(m)} \sim \mathcal{CN} (0,1)$ are complex gains which are i.i.d. in $l$ and $p$, and $\tilde{\theta}_{l,p}^{(m)}$ are azimuth angles of rays which are uniformly placed in the interval $(\theta_{p}^{(m)} - \Delta_{p}^{(m)}/2 , \theta_{p}^{(m)} + \Delta_{p}^{(m)}/2)$. Therefore, a cluster has mean AoA $\theta_{p}^{(m)}$ and AS $\Delta_{p}^{(m)}$. As L goes to infinity, the channel model yields the cluster CCM $\bm{R}_p^{(m)}\triangleq \mathbb{E} \{ \bm{h}_{p}^{(m)} (\bm{h}_{p}^{(m)})^{H} \}$ as
\begin{equation} \label{eq_CCM_integral}
    \bm{R}_p^{(m)} = \frac{1}{\Delta_{p}^{(m)}} \int_{\theta_{p}^{(m)} - \Delta_{p}^{(m)}/2}^{\theta_{p}^{(m)} + \Delta_{p}^{(m)}/2} \bm{a}(\theta)\bm{a}^{H}(\theta) d \theta
\end{equation}
at the $p$\textsuperscript{th} FT-CPI. Therefore, from \eqref{eq_rec_signal}, the total covariance matrix $\bm{\Psi}_p \triangleq \mathbb{E} \{ \bm{y}_{n,p} \bm{y}_{n,p}^{H} \}$ is expressed as
\begin{equation}
    \bm{\Psi}_p = \sum_{m=1}^{M} E^{(m)} \bm{R}_p^{(m)} + N_0 \bm{I}_N.
\end{equation}

\subsection{Time Variation Model} \label{sec_time_var}
The block-fading model \cite{goldsmith} is adopted to model the time variation in the channel. The channel $\bm{h}_{p}^{(m)}$ is assumed to be invariant within FT-CPIs of $T_{\text{F}}$ seconds, which consists of $N_{\text{F}}+N_{\text{S}}$ symbols as seen in Fig.~\ref{fig_cpi}. The channel varies after each FT-CPI as $p$ progresses. 

In the channel definition given in \eqref{eq_channel}, complex gains $\alpha_{l,p}^{(m)}$ are \emph{FT parameters},
 and mean AoA $\theta_{p}^{(m)}$ and AS $\Delta_{p}^{(m)}$ are \emph{ST parameters}. Time variation is mainly governed by the uncorrelatedness of $\alpha_{l,p}^{(m)}$ in FT-CPI index $p$, while ST parameters also vary through $p$ in a slower and correlated manner.

 In our work, variation of $\theta_p^{(m)}$ is modeled via a linear Gaussian state-space model while $\Delta_p^{(m)}$ is assumed to be constant. Transmitters are considered to move on a ring centered at the BS. $\theta_p^{(m)}$ and $\omega_p^{(m)}$ being the angular position and velocity in azimuth, respectively; the state vector in the $p$\textsuperscript{th} FT-CPI is denoted by $\bm{x}_p^{(m)} \triangleq [ \begin{matrix} \theta_p^{(m)} & \omega_p^{(m)} \end{matrix} ]^T$. The state equation is
\begin{equation}
\bm{x}_{p+1}^{(m)} = \bm{A} \bm{x}_{p}^{(m)} + \bm{\nu}_p^{(m)}, \qquad \bm{A}= \left[ \begin{matrix} 1 & T_{\text{F}} \\ 0 & 1 \end{matrix} \right],
\end{equation}
where $\bm{\nu}_p^{(m)} \sim \mathcal{N}(\bm{0},\bm{\Sigma}_{\nu})$ is the innovation with $\bm{\Sigma}_{\nu} = \text{diag} \{ \sigma_{\theta}^2, \sigma_{\omega}^2 \}$. The system initializes with a given $\bm{x}_0^{(m)}$.

\subsection{Practical Two-Stage Training Modes for Time-Varying Massive MIMO} \label{sec_2stageTraining}
The proposed structure of training and data transmission phases are illustrated in Fig.~\ref{fig_cpi}. FT-CPI and ST-CPI are defined to express the coherent interval in terms of FT and ST parameters. FT-CPIs consist of $N_{\text{F}}+N_{\text{S}}$ symbols, whereas ST-CPIs are longer and take $P$ FT-CPIs. In literature, they are also called gain and angle coherence time \cite{AngleGainCT}. They determine the repetition time for the training modes \emph{FT training} and \emph{ST beam tracking} as described in Section \ref{sec_est}, where beam tracking is performed more rarely. CPIs also determine the update times of different system blocks as described in Section \ref{sec_two_stage_bf}, where statistical blocks are updated more rarely.

This two-stage structure reduces the average computational complexity of beam tracking per time and enables low-complexity techniques for FT training with the aid of approximately known AoA, which is called \emph{beam-awareness}. However, the design parameters should be chosen such that the duration for $N_{\text{S}}+N_{\text{F}}$ symbols and $P$ FT-CPIs are shorter than the actual FT-CPI and ST-CPI, respectively. As a design guideline, we provide these intervals in Table \ref{tab_cpis} for different hardware settings, where $f_c$, $W$, $v$, $d$, and $\lambda$ are carrier frequency, bandwidth, speed, distance, and wavelength, respectively. 

\begin{table}[tb]
\caption{\textsc{FT-CPI and ST-CPI}}
	\centering
	\resizebox{0.48\textwidth}{!}{%
  \bgroup
\def\arraystretch{1.1}
	\begin{tabular}{|c||c|c|c|}
		\hline
		\multicolumn{4}{|c|}{Number of Symbols in an FT-CPI}
		\\
		\hline
		 & \multicolumn{3}{|c|}{Speed $v$ (m/s)}
		\\
		\hline
		$f_c/W$ & 0.1 & 1 & 10
		\\ \hline \hline
		30 & 10e6 & 1e6 & 100e3
		\\ \hline
		100 & 3e6 & 300e3 & 30e3 
		\\ \hline
		300 & 1e6 & 100e3 & 10e3
		\\ \hline
		1000 & 300e3 & 30e3 & 3e3
		\\ \hline
	\end{tabular}
	\begin{tabular}{|c||c|c|c|}
	    \hline
		\multicolumn{4}{|c|}{Number of FT-CPIs in an ST-CPI}
		\\
		\hline
		 & \multicolumn{3}{|c|}{$N$}
		\\
		\hline
		$d/\lambda$ & 16 & 64 & 128
		\\ \hline \hline
		1e3 & 1250 & 313 & 156
		\\ \hline
		3e3 & 3750 & 938 & 470
		\\ \hline
		10e3 & 12.5e3 & 3.1e3 & 1.6e3
		\\ \hline
		30e3 & 37.5e3 & 9.4e3 & 4.7e3
		\\ \hline
	\end{tabular}
 \egroup
	}
	\label{tab_cpis}			
\end{table}

FT-CPI, which corresponds to the channel coherence time \cite{goldsmith}, can be approximated as $0.1 /f_D$, where $f_D=vf_c/c$ is the Doppler frequency considering the radial movement, and $c$ is the speed of light. Then, $0.1 W /f_D$ gives the number of symbols in an FT-CPI. Therefore, there are approximately $c / (10v f_c/W)$ symbols in an FT-CPI. The time duration for an ST-CPI can be approximated as the beamwidth divided by angular speed. Beamwidth can be approximated as $2/N$ radians from $sin(\phi_{BW})\pi = 2\pi/N$. Angular speed can be approximated as $v/d$ radians per second considering the tangential movement where $d$ is the distance between the BS and the receiver. Therefore, the number of symbols in an ST-CPI is $2dW/(Nv)$. Furthermore, the number of FT-CPIs in an ST-CPI, which is the limit for $P$, is $20 (f_c d/c)/N$. Here, the ratio $f_c d/c=d/\lambda$ can be counted as a measure of required receive or transmit power, since it is directly related to the path-loss according to Friis transmission equation \cite{goldsmith}. Therefore, a selection of $d/\lambda$ ratio, for example 10e3, represents a group of practices with similar power requirements, for example 100 meters and 30GHz, or 1 kilometer and 3 GHz.

\section{General System Structure} \label{sec_two_stage_bf}
\begin{figure}[tb]
\centering
\resizebox{0.48\textwidth}{!}{%
\begin{tikzpicture}[>=triangle 45]
\def\xsc{0.9}
\def\ysc{1}
\def\xz{0*\xsc} 
\def\xa{\xz + 1.1*\xsc} 
\def\xb{\xa + 1.5*\xsc} %
\def\xc{\xb + 1.8*\xsc} 
\def\xd{\xc + 1.5*\xsc}
\def\xe{\xd + 1.7*\xsc}
\def\xf{\xe + 2*\xsc} 
\def\xg{\xf + 1.2*\xsc} 
\def\xh{\xg + 1.3*\xsc} 
\def\ya{0}
\def\yb{\ya - 0.6*\ysc} 
\def\yc{\yb - 1.5*\ysc} %
\def\yd{\yc - 1.3*\ysc} 
\def\ye{\yd - 1.5*\ysc} %
\def\yf{\ye - 1.3*\ysc} 
\def\yest{1.2*\ysc}
 
\node (n11) at (\xa,\yb)  {}; 
\node (n12) at (\xb,\yb) [coordinate] {};   
\node (n13) at (\xc,\yb) [block] {$\bm{W}^{(1)}$}; 
\node (n13t) at (n13.north) [anchor=south] {DBF};
\node (n14) at (\xd,\yb) [above]{$\bm{z}_{n,p}^{(1)}$}; 
\node (n15) at (\xe,\yb) [block] {$\hat{\tilde{\bm{h}}}_p^{(1,1)}$};
\node (n15t) at (n15.north) [anchor=south] {ICS-CMF};
\node (n16) [above] at (\xf,\yb) {$\hat{s}_{n,p}^{(1)}$};
\node (n1e) at ($(n14)-(0,\yest)$) [block] {Estimation for the 1\textsuperscript{st} Cluster};

\node (n23) at (\xc,\yc) {$\vdots$};
\node (n25) at (\xe,\yc) {$\vdots$};

\node (n30) at (\xz,\yc) {$\bm{y}_{n,p}$}; 
\node (n31) at (\xa,\yd) [block, minimum height=15em] {$\bm{S}$};
\node() at (n31.north) [anchor=south] {ABF};
\node (n32) at (\xb,\yd) [coordinate] {}; 
\node () at ($(n31.east)!0.5!(n32)$) [anchor=south] {$\bm{r}_{n,p}$};
\node (n33) at (\xc,\yd) [block] {$\bm{W}^{(m)}$}; 
\node (n33t) at (n33.north) [anchor=south] {DBF};
\node (n34) at (\xd,\yd) [above]{$\bm{z}_{n,p}^{(m)}$}; 
\node (n35) at (\xe,\yd) [block] {$\hat{\tilde{\bm{h}}}_p^{(m,m)}$};
\node (n35t) at (n35.north) [anchor=south] {ICS-CMF};
\node (n36) at (\xf,\yd) [above] {$\hat{s}_{n,p}^{(m)}$}; 
\node (n37) at (\xg,\yd) [block, minimum height=20em] {}; 
\node () at (n37) [rotate=90] {Digital Cluster Combiner}; 
\node (n3e) at ($(n34)-(0,\yest)$) [block] {Estimation for the m\textsuperscript{th} Cluster};

\node (n43) at (\xc,\ye) {$\vdots$};  
\node (n45) at (\xe,\ye) {$\vdots$};
\node (n51) at (\xa,\yf) [coordinate] {}; 
\node (n52) at (\xb,\yf) [coordinate] {};  
\node (n53) at (\xc,\yf) [block] {$\bm{W}^{(M)}$};
\node (n53t) at (n53.north) [anchor=south] {DBF};
\node (n54) at (\xd,\yf) [above]{$\bm{z}_{n,p}^{(M)}$}; 
\node (n55) at (\xe,\yf) [block] {$\hat{\tilde{\bm{h}}}_p^{(M,M)}$};
\node (n55t) at (n55.north) [anchor=south] {ICS-CMF};
\node (n56) at (\xf,\yf) [above] {$\hat{s}_{n,p}^{(M)}$};
\node (n5e) at ($(n54)-(0,\yest)$) [block] {Estimation for the M\textsuperscript{th} Cluster};

\coordinate (n17i) at (n11-|n37.west) {};
\coordinate (n57i) at (n51-|n37.west) {};
\draw[->] (n30) |- (n31);
\draw[->] (n31) -- (n33);
\draw[->] (n33) -- (n35);
\draw[->] (n35) -- (n37);

\draw[->] (n32) |- (n13);
\draw[->] (n13) -- (n15);
\draw[->] (n15) -- (n17i);

\draw[->] (n32) |- (n53);
\draw[->] (n53) -- (n55);
\draw[->] (n55) -- (n57i);

\draw[->,dashed] (n14) -- (n1e);
\draw[->,dashed] (n34) -- (n3e);
\draw[->,dashed] (n54) -- (n5e);

\coordinate (pb_l) at ($(n33.west)-(0.8*\xsc,0)$) {};
\coordinate (pb_r) at ($(n35.east)+(0.6*\xsc,0)$) {};
\draw[gray, very thick] (pb_l|-n15t.north) -- (pb_r|-n15t.north) -- ($(pb_r|-n1e.south)-(0,0.1*\ysc)$)-- ($(pb_l|-n1e.south)-(0,0.1*\ysc)$) -- (pb_l|-n15t.north);
\draw[gray, very thick] (pb_l|-n35t.north) -- (pb_r|-n35t.north) -- ($(pb_r|-n3e.south)-(0,0.1*\ysc)$)-- ($(pb_l|-n3e.south)-(0,0.1*\ysc)$) -- (pb_l|-n35t.north);
\draw[gray, very thick] (pb_l|-n55t.north) -- (pb_r|-n55t.north) -- ($(pb_r|-n5e.south)-(0,0.1*\ysc)$)-- ($(pb_l|-n5e.south)-(0,0.1*\ysc)$) -- (pb_l|-n55t.north);

\def\ybb{\yd + 2*\ysc}
\def\ycc{\yd + 1*\ysc}
\def\yee{\yd - 1*\ysc}
\def\yff{\yd - 2*\ysc}

\node (n18) at (\xh,\ybb) {$\hat{b}_{n,p}^{(1)}$};
\node (n28) at (\xh,\ycc) {$\vdots$};
\node (n38) at (\xh,\yd) {$\hat{b}_{n,p}^{(u)}$};
\node (n48) at (\xh,\yee) {$\vdots$};
\node (n58) at (\xh,\yff) {$\hat{b}_{n,p}^{(U)}$};
\coordinate (n17o) at (n18-|n37.east) {};
\coordinate (n37o) at (n38-|n37.east) {};
\coordinate (n57o) at (n58-|n37.east) {};
\draw[->] (n17o) -- (n18);
\draw[->] (n57o) -- (n58);
\draw[->] (n37o) -- (n38);
\end{tikzpicture}
}
\caption{Beamforming structure}
\label{fig_bf}
\end{figure}
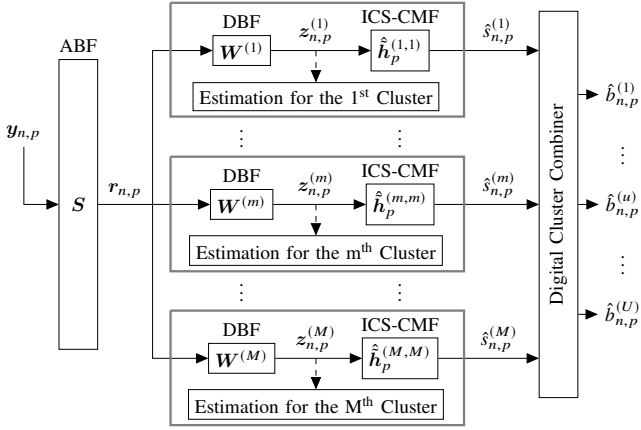

In this section, the general structure of the beamforming and processing system will be introduced, leaving the details of design procedures to Sections \ref{sec_est} and \ref{sec_geb}. The HBF structure is adopted as depicted in Fig.~\ref{fig_bf} for its practicality and energy efficiency. ABF produces inputs for $R$ RFCs via analog phase shifters and combiners, selecting all the interested clusters in the angular domain. 

After the ABF, different than conventional systems, statistical beamforming is employed in cluster-specific DBF blocks via RD-GEB in Section \ref{sec_geb}. DBFs create cluster-specific processing blocks, each of which includes an ICS-CMF and a per-cluster estimation block. The DBF is responsible for the formation of a cluster-specific subspace that mitigates the multi-cluster interference. Therefore, ICS-CMF and estimators are designed by ignoring the multiuser interference, yielding low complexity. ICS-CMF combines the outputs from the DBF as a matched filter. The chain of ABF, DBF, and ICS-CMF gradually reduces the signal dimension from $N$ to $R/M$, $D_m$, and 1, where $N > R/M \ge D_m \ge 1$. The output of DBF is also the place where the per-cluster estimation is performed, which consists of FT training and ST beam tracking, which will be detailed in Section \ref{sec_est}. 

Gains and phases of the outputs for clusters are corrected, and they are also combined by the digital cluster combiner (DCC) to reach symbol estimates for users, in the presence of multipath components ($U<M$) as described by \eqref{eq_cluster2user}. In this case, DCC is the block that brings the suitability for frequency-selective channels to the system.

In this structure, ABF and DBF are updated after each ST beam tracking rarely, while ICS-CMF and DCC are updated after each FT training more frequently.  

\subsection{DFT-Based Analog Beamformer for Full Cluster Coverage} \label{sec_abf}
The analog beamforming matrix is denoted by $\bm{S} \in \mathbb{C}^{N \times R}$ for $R$ RFCs, whose columns are selected among DFT bases which results in a practical implementation via analog phase shifters. Also, the columns are orthonormal such that $\bm{S}^{H} \bm{S} = \bm{I}_R$. The output of the ABF is
\begin{equation} \label{eq_rec_signal_after_abf}
    \bm{r}_{n,p} \triangleq \bm{S}^{H} \bm{y}_{n,p} = \sum_{m=1}^{M} \sqrt{E^{(m)}} \bar{\bm{h}}_{p}^{(m)} s_{n,p}^{(m)} + \bar{\bm{\eta}}_{n,p}
\end{equation}
where $\bar{\bm{h}}_{p}^{(m)} \triangleq \bm{S}^{H} \bm{h}_{p}^{(m)}$ and $\bar{\bm{\eta}}_{n,p} \triangleq \bm{S}^{H} \bm{\eta}_{n,p}$. After the analog stage, covariance matrices $\bm{R}_p^{(m)}$ and $\bm{\Psi}_p$ are modified as 
\begin{gather} \label{eq_cluster_ccm_after_abf}
    \bar{\bm{R}}_p^{(m)} = \bm{S}^{H} \bm{R}_p^{(m)} \bm{S},
\\ \label{eq_total_ccm_after_abf}
    \bar{\bm{\Psi}}_p = \bm{S}^{H} \bm{\Psi}_p \bm{S} = \sum_{m=1}^{M} E^{(m)} \bar{\bm{R}}_p^{(m)} + N_0 \bm{I}_R,
\end{gather}
where $\bar{\bm{\Psi}}_p \triangleq \mathbb{E} \{ \bm{r}_{n,p} \bm{r}_{n,p}^{H} \}$ and $\bar{\bm{R}}_p^{(m)} \triangleq \mathbb{E} \{ \bar{\bm{h}}_{p}^{(m)} (\bar{\bm{h}}_{p}^{(m)})^{H} \}$.

\subsection{Statistical Digital Beamformer for Inter-Cluster Separation}
The digital stage starts with a bank of DBFs. DBFs are cluster-specific digital combiners for which statistical beamforming is employed. The DBF for the $m$\textsuperscript{th} cluster $\bm{W}^{(m)} \in \mathbb{C}^{R \times D_m}$ has orthonormal columns such that $(\bm{W}^{(m)})^{H} \bm{W}^{(m)} = \bm{I}_{D_m}$. It processes the ABF output $\bm{r}_{n,p}$ and outputs $\bm{z}_{n,p}^{(m)} \triangleq \left(\bm{W}^{(m)}\right)^{H} \bm{r}_{n,p}$. Defining the total beamforming matrix $\bm{T}^{(m)} \triangleq \bm{S} \bm{W}^{(m)}$, the DBF output is
\begin{equation} \label{eq_z}
    \bm{z}_{n,p}^{(m)}  = (\bm{T}^{(m)})^{H} \bm{y}_{n,p} = \sum_{m'=1}^{M} \sqrt{E^{(m')}} \tilde{\bm{h}}_{p}^{(m,m')} s_{n,p}^{(m')} + \tilde{\bm{\eta}}_{n,p}^{(m)}
\end{equation}
where $\tilde{\bm{h}}_{p}^{(m,m')} \triangleq (\bm{T}^{(m)})^{H} \bm{h}_{p}^{(m')}$ is the IEC and $\tilde{\bm{\eta}}_{n,p}^{(m)} \triangleq (\bm{T}^{(m)})^{H} \bm{\eta}_{n,p} \sim \mathcal{CN}(\bm{0},N_0 \bm{I}_{D_m})$. After the digital stage, covariance matrices of $\tilde{\bm{h}}_p^{(m,m')}$ and $\bm{z}_{n,p}^{(m)}$, denoted by $\tilde{\bm{R}}_p^{(m,m')}$ and $\tilde{\bm{\Psi}}_p^{(m)}$, respectively, are given in \eqref{eq_R_tilde} and \eqref{eq_Psi_tilde}.
\begin{gather} \label{eq_R_tilde}
    \tilde{\bm{R}}_p^{(m,m')} = (\bm{T}^{(m)})^{H} \bm{R}_p^{(m')} \bm{T}^{(m)}
\\ \label{eq_Psi_tilde}
    \tilde{\bm{\Psi}}_p^{(m)} \hspace{-2pt} = (\bm{T}^{(m)})^{H} \bm{\Psi}_p \bm{T}^{(m)} \hspace{-2pt} = \hspace{-2pt} \sum_{m'=1}^{M} E^{(m')} \tilde{\bm{R}}_p^{(m,m')} \hspace{-2pt} + N_0 \bm{I}_{D_m} 
\end{gather}

\subsection{Intra-Cluster Spatial Channel Matched Filtering (ICS-CMF)}
Different from temporal channel matched filtering, ICS-CMF compensates for the indifference of statistically constructed slow-time DBF to fast-time variations in the channel. After beamforming, $D_m$ outputs are obtained for the $m$\textsuperscript{th} cluster. They are efficiently combined using the matched filtering method assuming the multi-cluster interference has already been suppressed to a negligible level by beamforming.\footnote{If multiple users ($\le D_m$) fall into the angular sector of the same cluster, joint intra-cluster processing can be applied \cite{Kurt19} both in data and training phases in RD subspace ($D_m$).} The output of ICS-CMF is obtained by
\begin{equation}
    \hat{s}_{n,p}^{(m)} \triangleq (\hat{\tilde{\bm{h}}}_p^{(m,m)})^{H} \bm{z}_{n,p}^{(m)}
\end{equation}
where $\hat{\tilde{\bm{h}}}_p^{(m,m)}$ is the estimate for the channel $\tilde{\bm{h}}_p^{(m,m)}$, which will be detailed in Section \ref{sec_est_ft}. The output is expressed as
\begin{equation} \label{eq_cluster_symbols}
    \hat{s}_{n,p}^{(m)} = \sum_{m'=1}^{M} \sqrt{E^{(m')}} (\hat{\tilde{\bm{h}}}_p^{(m,m)})^{H} \tilde{\bm{h}}_p^{(m,m')} s_{n,p}^{(m')} + (\hat{\tilde{\bm{h}}}_p^{(m,m)})^{H} \tilde{\bm{\eta}}_{n,p}^{(m)}
\end{equation}

\subsection{Digital Cluster Combiner (DCC)}
After ICS-CMF, the system has symbol estimates from different signal clusters as shown in \eqref{eq_cluster_symbols}. These symbol estimates have different gains and different delays. Furthermore, they might be multipath components from the same user, as described by \eqref{eq_cluster2user}. By substituting \eqref{eq_cluster2user} into \eqref{eq_cluster_symbols}, we have 
\begin{equation}
\begin{aligned}
    \hat{s}_{n,p}^{(m)} = & \sum_{m'=1}^{M} \sqrt{E^{(m')}} (\hat{\tilde{\bm{h}}}_p^{(m,m)})^{H} \tilde{\bm{h}}_p^{(m,m')} b_{n-l_{m'},p}^{(\mathcal{U}(m'))} 
    \\ & + (\hat{\tilde{\bm{h}}}_p^{(m,m)})^{H} \tilde{\bm{\eta}}_{n,p}^{(m)}
\end{aligned}
\end{equation}

To reach an estimate for the symbols $b_{n,p}^{(u)}$ from different users; delay, magnitude, and phases of symbol estimates from each cluster are aligned as a first step as
\begin{equation}
    \tilde{b}_{n,p}^{(m)} \triangleq \frac{\hat{s}_{n+l_m,p}^{(m)}}{\sqrt{E^{(m)}} (\hat{\tilde{\bm{h}}}_p^{(m,m)})^{H} \hat{\tilde{\bm{h}}}_p^{(m,m)}}
\end{equation}
which can be approximated as $\tilde{b}_{n,p}^{(m)} \cong b_{n,p}^{(\mathcal{U}(m))} + e_{n,p}^{(m)}$. Then, if the channel includes multipath components, implying $U<M$, symbol estimates for $M$ clusters are combined into symbol estimates for $U$ users, as
\begin{equation}
    \hat{b}_{n,p}^{(u)} \triangleq \sum_{m=1}^{M} p_{m}^{(u)} \, \tilde{b}_{n,p}^{(m)} \cong \left( \sum_{m=1}^{M} p_{m}^{(u)} \right) b_{n,p}^{(u)} + \sum_{m=1}^{M} p_{m}^{(u)} e_{n,p}^{(m)}
\end{equation}
for $u=1,\dots,U$, where $p_{m}^{(u)}$ is the combiner, which can be designed via various techniques, such as BLUE, maximal ratio combining, or selection combining. Note that the structure of $p_{m}^{(u)}$ is very simple such that $p_{m}^{(u)} \neq 0$ only for $m$ values for which $\mathcal{U}(m)=u$. 

The digital cluster combiner is introduced to have an inclusive conceptual design. With this stage, the proposed scheme gains generality over frequency-selective and flat-fading channels. Also, it enables the per-cluster perspective of the proposed system design. After the advantages of the per-cluster scheme are exploited, per-user outputs are obtained via DCC in a simple way. 

The design of $p_{m}^{(u)}$ is left out of scope, and its performance will not be evaluated since it requires a much more detailed scenario description. However, the quality of the symbol estimates at its input, which will be evaluated via SINR that will be defined in Section \ref{sec_perf_meas}, determines its performance directly in any case.

\section{Two-Stage Parameter Estimation Procedure} \label{sec_est}
The task of channel estimation is held in two stages in the proposed system. They are named FT training and ST beam-tracking, which are introduced in Section \ref{sec_2stageTraining}. To recite, there is an FT/ST separation in channel parameters, CPIs, estimation phases, and processing blocks in terms of their update rate. 

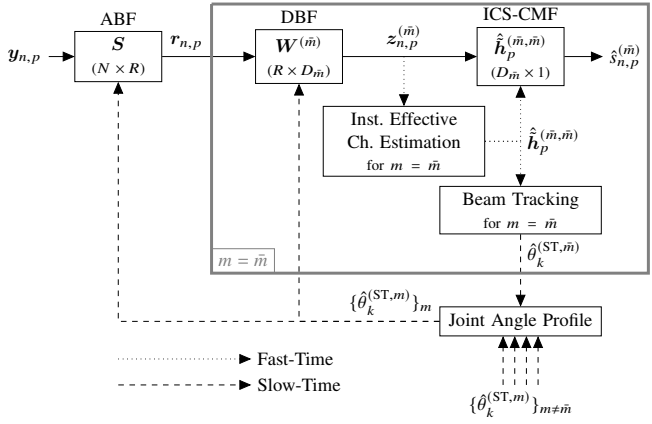
\begin{figure}[tb]
\centering
\resizebox{0.48\textwidth}{!}{%
\begin{tikzpicture}[>=triangle 45]
\def\xsc{1.1}
\def\ysc{1}
\def\xz{0*\xsc} 
\def\xa{\xz + 1.6*\xsc} 
\def\xb{\xa + 1.6*\xsc} %
\def\xc{\xb + 1.5*\xsc} 
\def\xd{\xc + 1.8*\xsc}
\def\xe{\xd + 2*\xsc}
\def\xf{\xe + 1.8*\xsc} 
\def\xg{\xf + 1*\xsc} 
\def\xh{\xg + 1.6*\xsc} 
\def\ya{0}
\def\yb{\ya - 0.6*\ysc} 
\def\yc{\yb - 1.7*\ysc} %
\def\yd{\yc - 1.3*\ysc} 
\def\ye{\yd - 1.6*\ysc} %
\def\yf{\ye - 1.3*\ysc} 
\def\yg{\yf - 2.1*\ysc} 
\def\yh{\yg - 1.2*\ysc} 
\def\yest{1.5*\ysc}

\node (n30) at (\xz,\yd) {$\bm{y}_{n,p}$}; 
\node (n31) at (\xa,\yd) [block, text width=4em] {$\bm{S}$ \\ \footnotesize ($N \times R$)};  
\node (n32) at (\xb,\yd) [coordinate] {}; 
\node () at ($(n31.east)!0.5!(n32)$) [anchor=south] {$\bm{r}_{n,p}$};
\node (n33) at (\xc,\yd) [block, text width=4em] {$\bm{W}^{(\bar{m})}$ \\ \footnotesize ($R \times D_{\bar{m}}$)};  
\node (n34) at (\xd,\yd) [above]{$\bm{z}_{n,p}^{(\bar{m})}$}; 
\node (n35) at (\xe,\yd) [block, text width=4em] {$\hat{\tilde{\bm{h}}}_p^{(\bar{m},\bar{m})}$ \\ \footnotesize ($D_{\bar{m}} \times 1$)};
\node (n36) at (\xf,\yd) [] {$\hat{s}_{n,p}^{(\bar{m})}$}; 

\node () at (n31.north) [anchor=south] {ABF};
\node (textdbf) at (n33.north) [anchor=south] {DBF};
\node () at (n35.north) [anchor=south] {ICS-CMF};

\draw[->] (n30) -- (n31);
\draw[->] (n31) -- (n33);
\draw[->] (n33) -- (n35);
\draw[->] (n35) -- (n36);

\def\xba{\xd - 2*\xsc}
\def\xbb{\xd - 2*\xsc}

\node (nfte) at (\xd,\ye) [block, text width=8em] {Inst. Effective \\ Ch. Estimation \\ \footnotesize for $m=\bar{m}$};
\node (nste) at (\xe,\yf) [block, text width=8em] {Beam Tracking \\ \footnotesize for $m=\bar{m}$};
\node (nprof) at (\xe,\yg) [block, text width=8em] {Joint Angle Profile};
\draw[->,dotted] (n34) -- (nfte);
\draw[->,dotted] (nfte) -| (n35);
\draw[->,dotted] (nfte) -| (nste);
\draw[->,dashed] (nste) -- (nprof);
\coordinate (other) at (\xe,\yh) {};
\draw[->,dashed] ($(other)-(0.3*\xsc,0)$) -- ($(nprof.south)-(0.3*\xsc,0)$);
\draw[->,dashed] ($(other)-(0.1*\xsc,0)$) -- ($(nprof.south)-(0.1*\xsc,0)$);
\draw[->,dashed] ($(other)+(0.1*\xsc,0)$) -- ($(nprof.south)+(0.1*\xsc,0)$);
\draw[->,dashed] ($(other)+(0.3*\xsc,0)$) -- ($(nprof.south)+(0.3*\xsc,0)$);

\draw[->,dashed] (nprof) -| (n31);
\draw[->,dashed] (nprof) -| (n33);

\node (otherset) at (other) [anchor=north] {$\{\hat{\theta}_k^{(\text{ST},m)}\}_{m\neq \bar{m}}$};

\node (textst) at (\xc,\yh) []  {Slow-Time};
\node (textft) at (textst.north west) [anchor=south west] {Fast-Time};
\draw[->,dotted] (n31|-textft) -- (textft.west);
\draw[->,dashed] (n31|-textst) -- (textst.west);
\node () at (nfte-|nste) [anchor=west] {$\hat{\tilde{\bm{h}}}_p^{(\bar{m},\bar{m})}$};
\node () at (nprof.west) [anchor=south east] {$\{\hat{\theta}_k^{(\text{ST},m)}\}_m$};
\node (textnsteout) at (nste.south) [anchor=north west] {$\hat{\theta}_k^{(\text{ST},\bar{m})}$};
\draw[gray,ultra thick] (n32|-textdbf.north) -- (n32|-textnsteout.south) -- (n36.east|-textnsteout.south) -- (n36.east|-textdbf.north) -- (n32|-textdbf.north);
\node () at (n32|-textnsteout.south) [anchor=south west, gray, block] {${m=\bar{m}}$};
\end{tikzpicture}
}
\caption{Estimation structure for the $\bar{m}$\textsuperscript{th} signal cluster}
\label{fig_bf2}
\end{figure}

The proposed \emph{per-cluster} estimation approach employs separate estimators for each cluster as seen in Fig.~\ref{fig_bf}. In Fig.~\ref{fig_bf2}, the estimation procedure is illustrated in more detail for one of the clusters. The fundamental aspect of the per-cluster estimation is that the estimation takes place after cluster-specific DBFs, which mitigate the interference better than conventional DFT beams. Also, the size of the effective channel decreases further. As a result, IEC estimators and beam trackers can be designed in a simpler way, as if only a single user exists, in the presence of multiple users or clusters.

On the other hand, in a conventional estimation scheme, the channel is estimated jointly after ABF, constructed with DFT beams. DFT beams are weak against interference, which charges the estimator with the interference mitigation task through pilots or other methods. Also, DBF depends on instantaneous channel estimates and needs more frequent updates, yet it is still responsible for interference mitigation for data transmission. Therefore, conventional methods suffer from performance loss, complexity increase, or a need for longer training overhead for channel estimation.

Beam trackers collect and use IEC estimates to estimate AoAs. It is very important that the channel estimator operates in fast-time and the beam tracker operates in slow-time, as shown by the dotted and dashed lines in Fig.~\ref{fig_bf2}. Therefore, ABF and DBF are updated in slow-time, while the ICS-CMF is updated in fast-time. Since DBF is slowly updated, a more complex design is tolerable which increases the interference mitigation performance.

\subsection{FT Estimation: Instantaneous Effective Channel (IEC) Estimation} \label{sec_est_ft}
The IEC estimator aims to estimate the rapidly changing (instantaneous) RD channel observed after DBF, which is the reason why it is called the \emph{effective} channel. Training sequences of length $N_\text{F}$ are transmitted, which are shown by the vector $\bm{s}^{(m)}=[s_{1+l_m,p}^{(m)} \cdots s_{N_{\text{F}}+l_m,p}^{(m)}]^{T}=[b_{1,p}^{(\mathcal{U}(m))} \cdots b_{N_{\text{F}},p}^{(\mathcal{U}(m))}]^{T}$, where $l_m$ is the cluster delay which is shown to be easy to track in Section \ref{sec_clusterdelay}. The received signals from $n=1+l_m$ to $n=N_{\text{F}}+l_m$ are collected and the techniques below are applied. In the subsequent sections, $l_m=0$ is assumed for the sake of simplicity.

\vspace{0.5em}
\subsubsection{Beam-Aware Least Squares (BA-LS) Estimation}
This technique observes the received signal after the ABF and DBF ($\bm{T}^{(m)}=\bm{S}\bm{W}^{(m)}$), which were set previously, to estimate the IEC denoted by $\tilde{\bm{h}}_p^{(m,m)}$. \emph{Beam-awareness} implies that the subspace created by $\bm{T}^{(m)}$ is cluster-specific and interference mitigation is accomplished before by the beamformer, which yields a simple design. This technique also does not need synchronization between the sources since it operates per-cluster. Therefore, $\bm{s}^{(m)}$ for different $m$'s are uncorrelated. 

The received signal $\bm{z}_{n,p}^{(m)}$ after $\bm{T}^{(m)}$ is as shown in \eqref{eq_z}. Received signals during the training are concatenated to obtain $\bm{z}_{:,p}^{(m)} \triangleq [(\bm{z}^{(m)}_{1,p})^{T} \dots (\bm{z}^{(m)}_{N_\text{F},p})^{T}]^{T}$, which is expressed as 
\begin{equation}
    \bm{z}_{:,p}^{(m)} = \sum_{m'=1}^{M} \sqrt{E^{(m')}} (\bm{s}^{(m')} \otimes \bm{I}_{D_m}) \tilde{\bm{h}}_p^{(m,m')} + \tilde{\bm{\eta}}_{:,p}^{(m)},
\end{equation}
where $\tilde{\bm{\eta}}_{:,p}^{(m)} \sim \mathcal{CN}\left(\bm{0}, \bm{I}_{N_{\text{F}}} \otimes (N_0 \bm{I}_{D_m}) \right)$, and $\otimes$ is the Kronecker product operator. Then, the received signal is processed by the LS estimator $\bm{Z}^{(m)}$ to obtain the LS estimate for the IEC $\tilde{\bm{h}}_p^{(m,m)}$ as 
\begin{equation} \label{eq_BA_LS}
    \hat{\tilde{\bm{h}}}_p^{(m,m)} \triangleq (\bm{Z}^{(m)})^{H} \bm{z}_{:,p}^{(m)}, \qquad \bm{Z}^{(m)} = \frac{(\bm{s}^{(m)} \otimes \bm{I}_{D_m})}{\sqrt{E^{(m)}} N_{\text{F}}}
\end{equation}
where the expression for $\bm{Z}^{(m)}$ is simplified with the assumption of $(\bm{s}^{(m)})^{H}\bm{s}^{(m)} = N_{\text{F}}$. Consequently, the channel estimate can be expressed as
\begin{equation} \label{eq_ch_est_exp}
    \hat{\tilde{\bm{h}}}_p^{(m,m)} = \tilde{\bm{h}}_p^{(m,m)} + \sum_{\substack{m'=1 \\ m' \neq m}}^{M} \frac{ \sqrt{E^{(m')}} ((\bm{s}^{(m)})^H \bm{s}^{(m')} )}{\sqrt{E^{(m)}} N_{\text{F}}} \tilde{\bm{h}}_p^{(m,m')} + \bm{\xi}_p^{(m)}
\end{equation}
where the second term is the inter-cluster interference and the last term is the error due to noise with $\bm{\xi}_p^{(m)} \sim \mathcal{CN} (\,\bm{0},\,\frac{N_0}{E^{(m)} N_{\text{F}}} \bm{I}_{D_m}\,)$.

\vspace{0.5em}
\subsubsection{Conventional Joint Estimation Techniques} \label{sec_conv_j_est}
In contrast to per-cluster BA-LS, a conventional estimation method might process the signals directly after the ABF and estimate the multi-cluster channels jointly. In this case, the $N_{\text{F}}R \times 1$ observation vector is $\bm{r}_{:,p} \triangleq \left[ \bm{r}_{1,p}^{T} \cdots \bm{r}_{N_{\text{F}},p}^{T} \right]^{T}$ and the $MR \times 1$ regressor vector would be
    $\bar{\bm{h}}_p^{(:)} \triangleq [ (\bar{\bm{h}}_p^{(1)})^{T} \cdots (\bar{\bm{h}}_p^{(M)})^{T} ]^{T}$.
The observation can be expressed as
\begin{equation}
    \bm{r}_{:,p} = \left( \left[ \sqrt{E^{(1)}} \bm{s}^{(1)} \, \dots \, \sqrt{E^{(M)}} \bm{s}^{(M)} \right] \otimes \bm{I}_R \right) \bar{\bm{h}}_p^{(:)} + \bar{\bm{\eta}}_{:,p}
\end{equation}
where $\bar{\bm{\eta}}_{:,p}$ is the AWGN with covariance $N_0 \bm{I}_{N_{\text{F}}R}$. With these definitions and the ones in Section \ref{sec_abf}, well-known LS and minimum mean squared error (MMSE) methods are implemented to estimate the concatenated channels $\bar{\bm{h}}_p^{(:)}$ jointly from the observation $\bm{r}_{:,p}$ after ABF \cite{kay}. The estimates are expressed as
\begin{align}
    \hat{\bar{\bm{h}}}_p^{(:),\text{LS}} & = \left(\bm{V}^{H} \bm{V} \right)^{-1} \bm{V}^{H} \bm{r}_{:,p} \label{eq_jls}
\\
    \hat{\bar{\bm{h}}}_p^{(:),\text{MMSE}} & = \bar{\bm{R}}_p^{(:)} \bm{V}^{H} \left(\bm{V} \bar{\bm{R}}_p^{(:)} \bm{V}^{H} + N_0 \bm{I}_{N_{\text{F}} R} \right)^{-1} \bm{r}_{:,p} \label{eq_jmmse}
\end{align}
where $\bm{V} \in \mathbb{C}^{(N_{\text{F}} R \times MR)}$ is defined so that $\bm{r}_{:,p} = \bm{V} \bar{\bm{h}}_p^{(:)} + \bar{\bm{\eta}}_{:,p}$, and $\bar{\bm{R}}_p^{(:)} \in \mathbb{C}^{(M R \times MR)}$ is a block diagonal matrix which comprises $\bar{\bm{R}}_p^{(m)}$ for $m~=~1,\dots,M$. Note that these joint methods require synchronization between users before the training phase.

\vspace{0.5em}
\subsubsection{Discussion on Cluster Delay and Delay Spread} \label{sec_clusterdelay}
The discrete delay of $m$\textsuperscript{th} cluster is denoted by $l_m$. The difference in delays of two clusters whose paths differ by $\Delta d$ is $\Delta l \triangleq |l_{m_1}-l_{m_2}| = \Delta d \times W / c$. For example, we have $\Delta l = 20$ for $W=100$ MHz and $\Delta d=60$ m. Assuming constant normalized bandwidth $W/f_c$, $d \times W$ becomes directly related to path loss from Friis transmission equation \cite{goldsmith}. Therefore, the range of a system with $W=10$ MHz increases compared to the one with $W=100$ MHz, and it could observe $\Delta d=600$ m, which results in $\Delta l = 20$ again. This difference in delays actually increases the effective delay spread. The joint techniques should observe the received signal starting from the first symbol of the cluster with the minimum delay until the last symbol of the cluster with the maximum delay. Therefore, the observation vector in Section \ref{sec_conv_j_est} might be much longer in reality. On the contrary, the per-cluster estimator BA-LS observes only one of the clusters at a time, which removes this problem. 

The estimation of the discrete delay is out of the scope of this paper. Once estimated, its tracking is straightforward. For example, for a source with speed $v$, consider its present position and that for $\Delta t$ later as two sources in the example above. Then, $\Delta d = v \Delta t$ in $\Delta l = \Delta d \times W / c$ expression. For $v=10$ m/s and $W=100$ MHz, $\Delta t$ should be 0.3 seconds to have $\Delta l = 1$. That is $l_m$ changes by 1 after 0.3 seconds, which is a very long period of time compared with an FT-CPI, or even an ST-CPI. It can be tracked as a very-slow-time parameter.

\subsection{ST Estimation: Beam Tracking}
In the ST beam tracking phase, ST parameters of the channel, namely AoA $\theta_{p}^{(m)}$ and AS $\Delta_{p}^{(m)}$, should be estimated, which determine the second-order statistics of the channel as seen in \eqref{eq_CCM_integral}. However, in this work, AS $\Delta_{p}^{(m)}$ is assumed to be constant and known.\footnote{AS estimation is studied in literature \cite{AngleGainCT} and it is out of the scope of this paper. It is shown in \cite{Kurt19} that the performance is not very sensitive to the exact value of AS.} AoA of each cluster is estimated separately by cluster-specific estimators by leveraging the IEC estimates from the latest $P$ FT-CPIs. Therefore, beam tracking is actually a calculation phase, and transmission of a new training sequence is not needed. After the beam tracking phase, beamformers are updated with the new information.

\vspace{0.5em}
\subsubsection{Assumed Slow-Time Model} \label{sec_beamtrack_st_model}
The overall design of ST operations, namely beam tracking and beamformer update, assume a \emph{coherent interval} in terms of ST parameters, namely \emph{ST-CPI}, which is longer than FT-CPIs as shown in Fig.~\ref{fig_cpi}. Therefore, these ST operations are repeated after a long time compared to FT operations. The length of ST-CPI is determined by the number of FT-CPIs within, denoted by $P$, as a design parameter, which should yield a similarity in terms of channel statistics such as
$\bm{R}_p^{(m)} \cong \bm{R}_{p+P}^{(m)}$. 

Due to the assumption of a different CPI, ST operations work under a mismatched channel model. Based on the aforementioned similarity, \emph{assumed} variables related to previously defined ones are needed, indicated by the superscript ST. The first example of ST variables is
\begin{equation}
    \bm{R}_k^{(\text{ST},m)} \cong \bm{R}_{p}^{(m)}, \qquad
    \text{for } p=(k-1)P+1, \dots , kP.
\end{equation}
As seen, ST variables are indexed by the ST-CPI index $k$, which is related to FT-CPI indices through $k=\lceil p/P \rceil$. Several ST variables will be used in the next sections without an explicit definition such as $\bm{\Psi}_k^{(\text{ST})}$, $\bar{\bm{R}}_k^{(\text{ST},m)}$, $\bar{\bm{\Psi}}_k^{(\text{ST})}$, $\bm{x}_k^{(\text{ST},m)}$, $\theta_k^{(\text{ST},m)}$, but they imply a similar relation.

The state-space model for angular variation defined in Section \ref{sec_time_var} is modified for ST operations as
\begin{align}
\bm{x}_{k+1}^{(\text{ST},m)}  & \triangleq \bm{A}^{(\text{ST})} \bm{x}_{k}^{(\text{ST},m)} + \bm{\nu}_k^{(\text{ST},m)},
\\
\theta_k^{(\text{ST},m)} & \triangleq [1 \,\,\, 0] \, \bm{x}_{k}^{(\text{ST},m)},
\end{align}
where $\bm{A}^{(\text{ST})} \triangleq \bm{A}^P$, $\bm{\nu}_k^{(\text{ST},m)} \sim \mathcal{N}(\bm{0}, \bm{\Sigma}_{\nu}^{(\text{ST})})$, and $\bm{\Sigma}_{\nu}^{(\text{ST})} = \sum_{i=1}^P \bm{A}^i \bm{\Sigma}_{\nu} (\bm{A}^i)^{T}$.

Furthermore, it is assumed that the overall beamformer structure has suppressed the multi-cluster interference to a negligible level. Therefore, the channel estimates expressed in \eqref{eq_ch_est_exp} are suboptimally assumed as
\begin{equation} \label{eq_assumed_ch_est}
    \hat{\tilde{\bm{h}}}_p^{(m,m)} \cong \tilde{\bm{h}}_p^{(m,m)} + \bm{\xi}_p^{(m)}
\end{equation}
with $\hat{\tilde{\bm{h}}}_p^{(m,m)} \sim \mathcal{CN} (\bm{0}, \bm{R}(\theta_k^{(\text{ST},m)}) + \frac{N_0}{E^{(m)} N_{\text{F}}} \bm{I}_{D_m})$ in the design of estimators. The parametric RD CCM $\bm{R}(\theta)$ is expressed as
\begin{equation} \label{eq_R_theta_original}
    \bm{R}(\theta) \triangleq (\bm{T}^{(m)})^{H} \bm{R}^{\text{FD}}(\theta) \bm{T}^{(m)},
\end{equation} 
where $\bm{R}^{\text{FD}}(\theta) \triangleq \frac{1}{\Delta} \int_{\theta - \Delta/2}^{\theta + \Delta/2} \bm{a}(\theta')\bm{a}^{H}(\theta') d \theta'$ is the full dimensional (FD) CCM for mean AoA $\theta$ and AS $\Delta$. \cite{Kurt20} shows that $\bm{R}^{\text{FD}}(\theta) \cong \, \text{diag}(\bm{a}(\theta)) \bm{D} \, \text{diag}(\bm{a}(\theta))^{H}$ where $(\, \bm{D} \,)_{a,b} = \text{sinc}\left((a-b)\cos(\theta)\sin(\Delta/2)\right)$ for $a,b=1,\dots,N$. Further inspired by \cite{Kurt20} for a simpler calculation of $\bm{R}(\theta)$, we can approximate it by
\begin{align} \label{eq_R_theta_simple}
    \bm{R}(\theta) & \cong \bm{E}(\theta) \bm{E}^{H}(\theta)
    \\
    \bm{E}(\theta) & \triangleq (\bm{T}^{(m)})^{H} \text{diag}(\bm{a}(\theta)) \bm{E}^{\text{FD}} \label{eq_R_theta_simple_E}
\end{align}
where $\bm{E}^{\text{FD}}$ has $\sqrt{\lambda_d}\bm{e}_d$ in its columns where $\lambda_d$ and $\bm{e}_d$ are eigenvalues and eigenvectors of $\bm{D}$. Noting $\bm{R}^{\text{FD}}(\theta)$ and $\bm{D}$ are effectively low-rank for practical $\Delta$ values, $\bm{E}^{\text{FD}}$ can be constructed with the most dominant $D_m$ eigenvalues and eigenvectors, yielding a size of $N \times D_m$ where $D_m \ll N$.\footnote{The number of columns of $\bm{E}^{\text{FD}}$ and the number of DFT outputs are equal since both are determined by the effective rank of the CCM.} $\bm{E}^{\text{FD}}$ is calculated and stored once, and $\bm{R}(\theta)$ can be calculated with $\mathcal{O}(N D_m^2)$ multiplications via \eqref{eq_R_theta_simple}, instead of $\mathcal{O}(N^2 D_m)$ via \eqref{eq_R_theta_original}, leveraging also the diagonal form of the $N \times N$ matrix $\text{diag}(\bm{a}(\theta))$ inside the $\bm{E}(\theta)$ expression.\footnote{$\bm{D}$ depends on $\theta$ through $\cos(\theta)\sin(\Delta/2)$, which is the transformed AS divided by $2\pi$, obtained from $\pi\sin(\theta+\Delta/2)-\pi\sin(\theta-\Delta/2)$ \cite{Kurt20}. \cite{Kurt19} shows the tolerance to AS errors, therefore dependency on $\theta$ can be neglected by selecting $\theta=0$. Nevertheless, one can store more than one $\bm{D}$ and $\bm{E}^{\text{FD}}$ matrices for quantized values of $\theta$ for a more accurate approximation.}

\vspace{0.5em}
\subsubsection{Beam-Aware Maximum-Likelihood (BA-ML) Estimator} \label{sec_ba-ml}
The observation vector used by the BA-ML estimator for the $m$\textsuperscript{th} cluster in the $k$\textsuperscript{th} ST beam tracking phase is
\begin{equation} \label{eq_obs_ML}
    \bm{f}_k^{(m)} \triangleq \left[ \, \begin{matrix} \left[\hat{\tilde{\bm{h}}}_{(k-1)P+1}^{(m,m)}\right]^{H}  \, \cdots  \, \left[\hat{\tilde{\bm{h}}}_{kP}^{(m,m)}\right]^{H}  \end{matrix}  \, \right]^{H}.
\end{equation}
 In order to parameterize it on $\theta$, we rewrite \eqref{eq_assumed_ch_est} as 
\begin{equation} \label{eq_assumed_ch_est_ML}
    \hat{\tilde{\bm{h}}}_p^{(m,m)} \cong \bm{E}(\theta_k^{(\text{ST},m)}) \bm{\beta}_p + \bm{\xi}_p^{(m)},
\end{equation}
where $\bm{\beta}_p \sim \mathcal{CN}(\bm{0},\bm{I}_{D_m})$ is the basis coefficient vector for the matrix $\bm{E}(\theta_k^{(\text{ST},m)})$ defined in \eqref{eq_R_theta_simple_E}, whose columns span the range space of $\bm{R}(\theta)$ in \eqref{eq_R_theta_original}. The observation in \eqref{eq_obs_ML} is parameterized on the sought unknown $\theta_k^{(\text{ST},m)}$ with the model in \eqref{eq_assumed_ch_est_ML}. However, $\bm{\beta}_p$ is also an unknown. In this case, the \emph{nonlinear least squares method}, which is equivalently the ML method for Gaussian cases \cite{stoica}, yields the estimated $\theta_k^{(\text{ST},m)}$ from the observation $\bm{f}_k^{(m)}$ as
\begin{equation} \label{eq_ba-ml_est_original}
    \hat{\theta}_{k+1}^{(\text{ST},m)} = arg \max_{\theta} \max_{\{\bm{\beta}_p\}} p(\bm{f}_k^{(m)} | \theta, \{\bm{\beta}_p\} ),
\end{equation}
which firstly finds and sets the maximizing set of $\{\bm{\beta}_p\}$ for each given $\theta$, then applies classical ML procedure on $\theta$. However, the given problem has $PD_m + 1$ unknowns but $PD_m$ equations. Therefore, we reduce the size of $\bm{\beta}_p$ to $D_m'<D_m$ and modify $\bm{E}(\theta_k^{(\text{ST},m)}) \in \mathbb{C}^{D_m \times D_m}$ as $\bm{E}'(\theta_k^{(\text{ST},m)}) \in \mathbb{C}^{D_m \times D_m'}$ by simply removing the weakest eigenvectors from $\bm{E}^{\text{FD}}$ in \eqref{eq_R_theta_simple_E}, which modifies \eqref{eq_assumed_ch_est_ML} as 
\begin{equation} \label{eq_BAML_ch_model}
    \hat{\tilde{\bm{h}}}_p^{(m,m)} \cong \bm{E}'(\theta_k^{(\text{ST},m)}) \bm{\beta}_p' + \bm{\xi}_p^{(m)}.
\end{equation}

Substituting \eqref{eq_BAML_ch_model} in \eqref{eq_obs_ML}, it is found in Appendix \ref{app_ml} that the AoA estimate can be calculated as
\begin{equation} \label{eq_BA_ML}
    \hat{\theta}_{k+1}^{(\text{ST},m)} = arg \min_\theta \text{tr} \left( \bm{M}(\theta) \left(\sum_{p=1}^{P}  \hat{\tilde{\bm{h}}}_p^{(m,m)} (\hat{\tilde{\bm{h}}}_p^{(m,m)})^{H}  \right) \right),
\end{equation}
where $\bm{M}(\theta) \triangleq \bm{I}_{D_m} - \bm{E}'(\theta) \left( \bm{E}'^{H}(\theta) \bm{E}'(\theta) \right)^{-1} \hspace{-3pt} \bm{E}'^{H}(\theta)$. 

\vspace{0.5em}
\subsubsection{Statistical Extended Kalman Filter (SEKF)}
A Bayesian method, particularly the Kalman filter, can exploit the correlation of AoA through time better due to slow variation. We propose the usage of second-order channel statistics as observation with EKF since it is a nonlinear function of AoA. The statistics are obtained via sample-mean covariance matrix, and the observation vector for EKF is obtained as
\begin{equation} \label{eq_SEKF_r}
    \bm{f}_k^{(m)}=\text{vec} \left\{ \frac{1}{P} \sum_{p=(k-1)P+1}^{kP} \hat{\tilde{\bm{h}}}_p^{(m,m)} (\hat{\tilde{\bm{h}}}_p^{(m,m)})^{H} \right\}
\end{equation}
where $\text{vec}\{\cdot\}$ is the vectorization operator. Assuming angular coherence, the observation can be rewritten as 
\begin{equation} \label{eq_SEKF_r_model}
    \bm{f}_k^{(m)}=\text{vec} \left\{ \bm{R}_f(\theta_k^{(\text{ST},m)}) \right\} + \bm{q}_k^{(m)}
\end{equation}
whose first term is the mean with $\bm{R}_f(\theta_k^{(\text{ST},m)}) \triangleq \bm{R}(\theta_k^{(\text{ST},m)}) + \frac{N_0}{E^{(m)} N_{\text{F}}} \bm{I}_{D_m}$, where \eqref{eq_R_theta_simple} can be used for $\bm{R}(\theta)$. The second term $\bm{q}_k^{(m)}$ is the zero-mean error vector with covariance $\bm{Q}_k^{(m)}$, which is found in Appendix \ref{app_sekf_Q} as
\begin{equation} \label{eq_SEKF_Q}
    \bm{Q}_k^{(m)} = \frac{1}{P} \left(\bm{R}_f (\theta_k^{(\text{ST},m)})\right)^{*} \otimes \bm{R}_f (\theta_k^{(\text{ST},m)})
\end{equation}

To implement EKF, $\text{vec} \{ \bm{R}_f(\theta_k^{(\text{ST},m)}) \}$ in \eqref{eq_SEKF_r_model} is linearized using the first-order Taylor expansion, which yields
\begin{equation} \label{eq_SEKF_r_model_linear1}
    \bm{f}_k^{(m)} \cong \bm{B}_k^{(m)} \bm{x}_k^{(\text{ST},m)}  + \bm{q}_k^{(m)} + \{known\,\&\,constant\}
\end{equation}
where $\bm{B}_k^{(m)}$ is the Jacobian matrix of $\text{vec} \{ \bm{R}_f(\theta) \}$ with respect to $\bm{x}_k^{(\text{ST},m)}$, to be calculated at $\hat{\bm{x}}_{k|k-1}^{(\text{ST},m)}=[\hat{\theta}_{k|k-1}^{(\text{ST},m)} \quad \hat{\omega}_{k|k-1}^{(\text{ST},m)}]$. It is found as
\begin{equation} \label{eq_SEKF_r_model_linear2}
    \bm{B}_k^{(m)} =\left[ \left( \frac{ \partial \text{vec} \{ \bm{R}(\theta) \}}{\partial \theta} \right) \Big|_{\theta = \hat{\theta}_{k|k-1}^{(\text{ST},m)}} \qquad \bm{0} \right]
\end{equation}
whose second column, derivative with respect to speed, is zero due to the angular coherence assumption. The derivative with respect to the angular position can be calculated numerically or analytically using the simple structure in \eqref{eq_R_theta_simple}.

Using these definitions, well-known iterations of EKF \cite{kay} are applied for each cluster $m$. These iterations are shown below for the $k$\textsuperscript{th} beam tracking phase.
\begin{equation} \label{eq_SEKF_K}
    \bm{K} =\bm{\Sigma}_{k|k-1}^{(\text{ST},m)} (\bm{B}_k^{(m)})^{H} \hspace{-2pt} \left( \bm{B}_k^{(m)} \bm{\Sigma}_{k|k-1}^{(\text{ST},m)} (\bm{B}_k^{(m)})^{H} \hspace{-2pt} + \bm{Q}_k^{(m)} \right)^{-1} \hspace{-6pt}
\end{equation}
\begin{align}
    \hat{\bm{x}}_{k|k}^{(\text{ST},m)} & = \hat{\bm{x}}_{k|k-1}^{(\text{ST},m)} + \bm{K} \left( \bm{f}_k^{(m)} - \text{vec}\{\bm{R}_f(\hat{\theta}_{k|k-1}^{(\text{ST},m)})\} \right)
\\
    \bm{\Sigma}_{k|k}^{(\text{ST},m)} & = \bm{\Sigma}_{k|k-1}^{(\text{ST},m)} - \bm{K} \bm{B}_k^{(m)} (\bm{\Sigma}_{k|k-1}^{(\text{ST},m)})^{H}
\\
    \hat{\bm{x}}_{k+1|k}^{(\text{ST},m)} & = \bm{A}^{(\text{ST})} \hat{\bm{x}}_{k|k}^{(\text{ST},m)}
\\
    \bm{\Sigma}_{k+1|k}^{(\text{ST},m)} & = \bm{A}^{(\text{ST})} \bm{\Sigma}_{k|k}^{(\text{ST},m)} (\bm{A}^{(\text{ST})})^{T} + \bm{\Sigma}_{\nu}^{(\text{ST})}
\end{align}

After the $k$\textsuperscript{th} ST beam tracking, beamformers are updated with predicted AoAs $\hat{\theta}_{k+1|k}^{(\text{ST},m)}$ for $m=1,\dots,M$, which are the first elements of $\hat{\bm{x}}_{k+1|k}^{(\text{ST},m)}$ for $m=1,\dots,M$.

\vspace{0.5em}
\subsubsection{Orthogonal Matching Pursuit (OMP)}
Previous methods work per-cluster in the subspace created by the DBF $\bm{W}^{(m)}$, under our proposed per-cluster estimation scheme. Alternatively, conventional methods might directly work on the output of the ABF $\bm{S}$. We adapt the well-known OMP technique to our framework to obtain this alternative. 

OMP is a compressed sensing method that solves the systems in the form of $\bm{f}=\bm{G} \bm{x}$, where $\bm{G}$ is a fat matrix. For angular estimation, $\bm{f}$ and $\bm{G}$ could be set as the received signal vector and a matrix with steering vectors at the angles to be searched in its columns, respectively. Then, the angular estimates could be the angles of the columns of $\bm{G}$ related to the dominant values in $\bm{x}$. However, this form could fail when the observation is noisy and weights are zero-mean random variables, that is, the channel is subject to fading. 

The aforementioned adaptation is about the training sequence usage, presence of a preprocessing (ABF), countermeasures to the near-far effect, and most importantly, the utilization of the previous $P$ estimates. The challenge in the last factor is that the channel gains are uncorrelated in the previous $P$ FT-CPIs, and careless integration of them might result in the weights averaging out to zero. 

For our adaptation, the system to be solved is 
$\bm{F} = \bm{G} \bm{X}$
where $\bm{F} \in \mathbb{C}^{RN_{\text{F}} \times P}$ includes all the observed signals at the FT-training phases from the previous $P$ FT-CPIs. For the $k$\textsuperscript{th} beam tracking phase, it can be written as
\begin{equation}
    \bm{F} = \left[ \bm{r}_{:,(k-1)P+1} \cdots \bm{r}_{:,kP} \right],
\end{equation}
where $\bm{r}_{:,p}$ is as given in Section \ref{sec_conv_j_est}.
The matrix $\bm{G} \in \mathbb{C}^{RN_{\text{F}} \times M N_{\theta}}$ is in the form of 
$\bm{G} \triangleq \left[ \bm{G}^{(1)} \cdots \bm{G}^{(M)} \right]$ with
\begin{equation} \label{eq_omp_a_m}
    \bm{G}^{(m)} \triangleq \sqrt{E^{(m)}} \left( \bm{s}^{(m)} \otimes \bm{S}^{H} \left[ \bm{a}(\theta_1^{(m)}) \cdots \bm{a}(\theta_{N_{\theta}}^{(m)})\right] \right),
\end{equation}
where $N_{\theta}$ is the number of angles to be searched per cluster. The $c$\textsuperscript{th} column of $\bm{G}$, denoted by $\bm{G}_{:c}$, is affiliated with a cluster and an AoA. Let the functions $M^{OMP}(c)$ and $\theta^{OMP}(c)$ map these columns to clusters and AoAs, respectively. With these definitions, the modified OMP algorithm is given in Algorithm \ref{algorithm_omp}.

\begin{algorithm}
\begin{algorithmic}[1]
\REQUIRE $\bm{F}$, $\bm{G}$, $M^{OMP}(\cdot)$, $\theta^{OMP}(\cdot)$
\STATE $\tilde{\bm{F}}=\bm{F}$
\STATE $\mathcal{M}=\{1,\cdots,M\}$
\STATE $i=0$
\WHILE{$\mathcal{M}\neq \{\}$}
\STATE $i \leftarrow i+1$
\STATE $c_i = arg\max_c \left|| \left( \bm{G}_{:c} \right)^{H} \tilde{\bm{F}} |\right|_2^2$
\STATE $\bar{m}=M^{OMP}(c_i)$
\IF {$\bar{m} \in \mathcal{M}$}
\STATE $\mathcal{M} \leftarrow ( \mathcal{M}-\{\bar{m}\} )$
\STATE $\hat\theta_{k+1}^{(\text{ST}, \bar{m})}=\theta^{OMP}(c_i)$
\ENDIF
\STATE $\tilde{\bm{G}}= \left[\bm{G}_{:c_1} \cdots \bm{G}_{:c_i} \right]$
\STATE $\tilde{\bm{F}}= \bm{F} - \tilde{\bm{G}} \left( \tilde{\bm{G}}^{H} \tilde{\bm{G}} \right)^{-1} \tilde{\bm{G}}^{H} \bm{F}$
\ENDWHILE
\end{algorithmic}
    \caption{Modified OMP}
    \label{algorithm_omp}
\end{algorithm}

Considering the system $\bm{F} = \bm{G} \bm{X}$ with these definitions, it is seen that the rows of the weight matrix $\bm{X} \in \mathbb{C}^{MN_{\theta} \times P}$ are actually an angular map of signal presence monitored throughout $P$ FT-CPIs. In this regard, it is expected to be \emph{row-sparse}, that is, only a few rows have significant nonzero content. Therefore, we adapt the selection mechanism of the columns of the matrix $\bm{G}$ in the conventional OMP method as in step 6 of the Algorithm to reflect the row-sparsity of the matrix $\bm{X}$. The proposed mechanism also solves the problem of coherency, where the channel gains are uncorrelated through FT-CPIs. In step 7, estimates are categorized according to cluster affiliation, which is a general problem in joint estimation techniques. The categorization is, in fact, strengthened by the difference of the training sequences as seen in \eqref{eq_omp_a_m}. Then, steps from 8 to 11 handle the multiple estimates from the same cluster, which is needed due to AS and near-far effect. Note that this method requires pilot transmission and therefore synchronization between users, unlike the proposed per-cluster methods BA-ML and SEKF.

\section{Statistical Beamformer Construction} \label{sec_geb}
\subsection{Full Dimensional GEB (FD-GEB)}
GEB is a statistical beamformer whose performance is widely evaluated in \cite{Kurt19}.
Similar to the well-known Capon beamformer \cite{stoica}, GEB minimizes interference while the intended signal power is kept constant. However, GEB accomplishes this task over variances as 
\begin{align} \label{eq_geb_problem}
\begin{aligned}
    & \min_{\bm{w}} && \bm{w}^{H} \bm{\Psi}_k^{(\text{ST})} \bm{w} & \text{subject to} && \bm{w}^{H} \bm{R}_k^{(\text{ST},m)} \bm{w} = c
\end{aligned}
\end{align}
where $\bm{w}$ is a beamformer vector, and the terms are variances of $\bm{w}^{H}\bm{y}_{n,p}$ and $\bm{w}^{H}\bm{h}_{p}^{(m)}$, respectively. The covariance matrices can be calculated parametrically, and $c$ is a constant. The solution to this problem requires generalized eigendecomposition of the matrix pair $(\bm{R}_k^{(\text{ST},m)} ,\bm{\Psi}_k^{(\text{ST})})$, expressed as 
\begin{equation}
    \bm{R}_k^{(\text{ST},m)} \bm{e} = \lambda \bm{\Psi}_k^{(\text{ST})} \bm{e}.
\end{equation}
Choosing $\bm{w}_{\text{opt}}=\bm{e}_{\text{max}}$ solves the problem in \eqref{eq_geb_problem} and yields the maximum signal-to-interference-plus-noise ratio (SINR), where the generalized eigenvector $\bm{e}_{\text{max}}$ corresponds to the maximum generalized eigenvalue $\lambda_{max}$. 

FD-GEB is a single-stage beamformer where $\bm{T}^{(m)} \in \mathbb{C}^{N \times D_m}$ is constructed with the most dominant $D_m$ generalized eigenvectors, using either a fully digital or a fully analog beamformer (with the double phase shifter structure).

\subsection{Reduced Dimensional GEB (RD-GEB)}
For HBFs, GEB can be implemented in the subspace created by an ABF $\bm{S}$ using the effective CCM pair $(\bar{\bm{R}}_k^{(\text{ST},m)},\bar{\bm{\Psi}}_k^{(\text{ST})})$, given in \eqref{eq_cluster_ccm_after_abf} and \eqref{eq_total_ccm_after_abf}, instead of full-dimensional CCM pair $(\bm{R}_k^{(\text{ST},m)},\bm{\Psi}_k^{(\text{ST})})$. However, we propose a slightly different technique that is more robust and integrated with the design of the ABF $\bm{S}$. After the AoA estimates $\hat{\theta}_k^{(\text{ST},m)}$ for $m=1,\dots,M$ are produced in $k-1$\textsuperscript{st} ST beam tracking phase, ABF is formed by the selection of the $R$ DFT bases which are directed to the closest angles to $\hat{\theta}_k^{(\text{ST},m)}$. The DFT frequencies are
\begin{equation}
    \phi_k = \frac{2\pi}{N} k, \quad k=1,\dots,N
\end{equation}
and the $k$\textsuperscript{th} DFT basis is $\bm{u}(\phi_k)$ where $\bm{u}(\phi) \triangleq \frac{1}{\sqrt{N}} [ \begin{matrix} 1 & e^{j\phi} & \dots & e^{j(N-1)\phi} \end{matrix}]^{T}$. Then, $R/M$ DFT frequencies are selected for each $m$ such that $\sum_{r=1}^{R/M} |\phi_{k_r^{(m)}} - \pi\sin(\hat{\theta}_k^{(\text{ST},m)})|$ is minimized, where $k_r^{(m)}$ are the indices for the selected set.\footnote{Note that $k_r^{(m)}$ are consecutive in $r$ for an arbitrary $m$, that is $k_r^{(m)}~=~\bar{k}^{(m)}+r$ for some $\bar{k}^{(m)}$. Therefore, the selection is straightforward.} Then the ABF $\bm{S}$ is constructed as
\begin{align}
    \bm{S} & = [ \begin{matrix} \tilde{\bm{S}}^{(1)} & \dots & \tilde{\bm{S}}^{(M)} \end{matrix}], \label{eq_abf_total}
\\ 
    \tilde{\bm{S}}^{(m)} & = [ \begin{matrix} \bm{u}(\phi_{k_1^{(m)}}) & \dots & \bm{u}(\phi_{k_{R/M}^{(m)}}) \end{matrix}].\label{eq_abf_cluster}
\end{align}

We assume that each possible DFT basis selects an angular sector, such that $\bm{u}(\phi_k)$ selects the angular sector from $\theta_1$ to $\theta_2$, where $\pi\sin(\theta_1)=\phi_k - \pi/N$ and $\pi\sin(\theta_2)=\phi_k + \pi/N$. These angular sectors form a codebook of covariance matrices $\{\bm{C}_k\}$ for $k=1,\dots,N$, where
\begin{equation}
    \bm{C}_k \triangleq \frac{N}{2 \pi} \int_{\phi_k - \pi/N}^{\phi_k + \pi/N} \bm{u}(\phi) \bm{u}^{H}(\phi) d\phi.
\end{equation}
Using this codebook, approximated CCMs for the clusters are calculated as
\begin{align}
    \hat{\bar{\bm{R}}}_k^{(\text{ST},m)} & = \bm{S}^{H} \left( \frac{M}{R} \sum_{r=1}^{R/M} \bm{C}_{k_r^{(m)}} \right) \bm{S}
\\
    \hat{\bar{\bm{\Psi}}}_k^{(\text{ST})} & = \sum_{m=1}^M E^{(m)} \hat{\bar{\bm{R}}}_k^{(\text{ST},m)} + N_0 \bm{I}_R
\end{align}

Then, the DBF $\bm{W}^{(m)}$ is obtained using the most dominant $D_m$ generalized eigenvectors of the matrix pair $(\hat{\bar{\bm{R}}}_k^{(\text{ST},m)}, \hat{\bar{\bm{\Psi}}}_k^{(\text{ST})})$, which yields the total beamformer as $\bm{T}^{(m)}=\bm{S}\bm{W}^{(m)}$. 

Columns of $\bm{W}^{(m)}$ (generalized eigenvectors) are not necessarily orthogonal, therefore they are orthogonalized via QR decomposition, assigning the obtained vectors as the new $\bm{W}^{(m)}$ such that $(\bm{W}^{(m)})^{H} \bm{W}^{(m)} = (\bm{T}^{(m)})^{H} \bm{T}^{(m)} = \bm{I}_{D_m}$. This process does not bring any information loss and the range space of $\bm{W}^{(m)}$ is preserved. In addition, it has less computational complexity than eigendecomposition, and it provides numerical stability and notational simplicity.

\section{Computational Complexity} \label{sec_complexity}
Computational complexities of all the alternatives are tabulated in Table \ref{tab_complex} in terms of the number of multiplications using $\mathcal{O}(\cdot)$ notation. The dominant operations resulting in these complexities are also given in the third column where MI, MM, and ED mean matrix inversion, matrix multiplication, and eigendecomposition, respectively. Also * means a size reduction in inversion is possible and applied in favor of related techniques. Proposed per-cluster operations BA-LS, BA-ML, SEKF, and GEB construction are additionally multiplied by $M$ since they are applied for each cluster separately. In addition, all beam tracking methods and GEB construction are applied once in every $P$ FT-CPIs. Instantaneous channel estimators include the training size $N_{\text{F}}$, and search-based beam trackers BA-ML and OMP include $N_{\theta}$, which is the number of AoAs to be searched per cluster. For joint LS and MMSE channel estimators, and the MMSE beamformer, reduced matrix inversion sizes are shown using properties of Kronecker product and Woodbury matrix identity, although they are originally larger. Multiplications with matrices expressed by a Kronecker product, especially when it includes an identity matrix, might also yield a reduced number of multiplications when inspected in detail, such as the multiplication in \eqref{eq_BA_LS}, which yields $\mathcal{O}(M N_{\text{F}} D_m)$ instead of $\mathcal{O}(M N_{\text{F}} D_m^2)$.

Joint LS channel estimation in \eqref{eq_jls} has the inverse of a matrix of size $MR \times MR$ and multiplication of matrices of sizes $MR \times MR$, $MR \times N_{\text{F}}R$, and $N_{\text{F}}R \times 1$. Joint MMSE channel estimation in \eqref{eq_jmmse} has the inverse of a matrix of size $N_{\text{F}}R \times N_{\text{F}}R$ and multiplication of matrices of sizes $MR \times N_{\text{F}}R$, $N_{\text{F}}R \times N_{\text{F}}R$, and $N_{\text{F}}R \times 1$. The size of the matrix inversion can be reduced to $M$ for Joint LS using a property of Kronecker product, and to $MR$ for Joint MMSE using the Woodbury matrix identity. On the other hand, MMSE BF construction in \eqref{eq_mmse_bf} is dominated by a matrix inversion of size $R \times R$ or $N \times N$, however, it can be reduced to an $M \times M$ inverse by the Woodbury matrix identity. Usage of these properties and identities changes also the placement of matrices and matrix multiplication strategy, and the reduced complexities in Table \ref{tab_complex} are obtained. For GEB, eigendecomposition for $x\times x$ matrices has a complexity of $\mathcal{O}(x^3)$, where this size is $N$ for FD-GEB and $R$ for RD-GEB. 

We can roughly assume that $M \cong D_m \cong R/M < N_\text{F} \ll N_\theta < N < P$ (see Table \ref{tab_parameters}). Accordingly, we can sort the computational complexity of techniques as BA-LS < Joint LS $\ll$ Joint MMSE for instantaneous channel estimation, SEKF < BA-ML $\ll$ OMP for beam tracking, and RD-GEB $\ll$ FD/RD MMSE BF < FD-GEB for beamformer construction.

Consequently, it is clearly seen that the proposed per-cluster estimation approach with statistical beamforming, represented by BA-LS, BA-ML, SEKF, and GEB, yields very low computational complexity compared with the conventional joint estimation approach with instantaneous beamforming, represented by joint LS, joint MMSE, OMP, and MMSE BF.

\begin{table}[tb]
\caption{\textsc{Average Computational Complexity Per FT-CPI}}
	\centering
	\resizebox{0.48\textwidth}{!}{%
  \bgroup
\def\arraystretch{1.2}
	\begin{tabular}{|c|c|c|c|}
		\hline
		\textbf{Task} & \textbf{Technique} & \textbf{\# of Multiplications} & \textbf{Dominant Op.}
		\\ \hline
		Inst. Eff. & BA-LS & $\mathcal{O}(M N_{\text{F}} D_m)$ & \eqref{eq_BA_LS}
		\\  \cline{2-4}
		Channel & Joint LS & $\mathcal{O}(M^3) + \mathcal{O}(M N_{\text{F}} R)$ & MI* \& MM 
		 \\  \cline{2-4}
		Estimation & Joint MMSE & $\mathcal{O}(M^3 R^3) + \mathcal{O}(M^2 N_{\text{F}} R^3)$ & MI* \& MM
		 \\  \hline
		& \multirow{2}{*}{BA-ML} & $\mathcal{O}(M N_{\theta} N D_m^2 / P)$ & \multirow{2}{*}{\eqref{eq_R_theta_simple_E}, \eqref{eq_BA_ML}}
		 \\ 
        \multirow{2}{*}{Beam} &  & $ + \mathcal{O}(M D_m^2 + M N_{\theta} D_m^3 / P)$ &
		 \\  \cline{2-4}
		\multirow{2}{*}{Tracking} & \multirow{2}{*}{SEKF} & $\mathcal{O}(M N D_m^2 / P)$ & 
        \multirow{2}{*}{\eqref{eq_R_theta_simple_E}, \eqref{eq_SEKF_r}, \eqref{eq_SEKF_K}*}
        \\ & & $ + \mathcal{O}(M D_m^2) + \mathcal{O}(M D_m^4 / P)$ &
		\\  \cline{2-4}
		& OMP & $\mathcal{O}(M N_{\theta} R N_{\text{F}})$ & Step 6 in Alg. \ref{algorithm_omp}
		 \\ \hline
		& FD-GEB & $\mathcal{O}(M N^3 / P)$ & ED
		 \\  \cline{2-4}
		Beamformer & RD-GEB & $\mathcal{O}(M R^3 / P)$ & ED
		 \\  \cline{2-4}
		Construction & FD-MMSE BF & $\mathcal{O}(M^3 + NM^2)$ & MI* \& MM
            \\ \cline{2-4}
		& RD-MMSE BF & $\mathcal{O}(M^3 + RM^2)$ & MI* \& MM
		 \\  \hline
	\end{tabular}
 \egroup
	}
	\label{tab_complex}			
\end{table}

\section{Numerical Evaluations}
\subsection{Auxiliary Definitions} \label{sec_aux_def}
For comparison purposes in numerical results, some auxiliary definitions are needed. Firstly, the prefix \emph{RD} (reduced dimensional) implies that the DBF is working after an ABF in an HBF system, whereas \emph{FD} (full dimensional, or fully digital) refers to a system where the beamformer is fully digital and there is no ABF, therefore $\bm{S}=\bm{I}_N$ and there are $R=N$ RFCs. 

\emph{DFT beamformer} has the same ABF $\bm{S}$ as RD-GEB. However, the DBF $\bm{W}^{(m)}$ is a matrix of ones and zeros, only selecting the RFCs (columns of $\bm{S}$) for the $m$\textsuperscript{th} cluster in \eqref{eq_abf_total} so that $\bm{T}^{(m)}=\bm{S}\bm{W}^{(m)}=\tilde{\bm{S}}^{(m)}$. This setting of DBF can be seen as the most primitive attempt to create cluster-specific subspaces after ABF. The improvement brought by RD-GEB can be seen from its difference from DFT BF. 

The \emph{periodogram} is a spectral estimation method \cite{stoica}, which is the implicit basis for a variety of techniques in the beam tracking literature.  It is applied in the range space of the DFT beamformer in the case of HBF structures. The peak of $\rho(\theta) \triangleq \sum_{n,p} |\bm{a}(\theta)^{H} \tilde{\bm{S}}^{(m)} \bm{z}_{n,p}^{(m)}|^2$ is searched for the $m$\textsuperscript{th} cluster, where $\bm{z}_{n,p}^{(m)}=(\tilde{\bm{S}}^{(m)})^{H} \bm{y}_{n,p}$. Inspecting \eqref{eq_abf_cluster}, note that $\rho(\theta)$ measures the power at the RFC outputs one by one at certain $\theta$ values. Therefore, it represents the related studies in Section \ref{sec_literature}. Considering all the training phases in an ST-CPI are used, the summation above is taken from $PN_\text{F}$ samples.

\emph{MMSE beamformer} is another beamforming method, different than MMSE channel estimator, for which an MMSE estimator \cite{kay} is constructed where symbols $s_{n,p}^{(m)}$ for $m=1,\dots,M$ are treated as the multiple unknown parameters with the observation $\bm{y}_{n,p}$ in \eqref{eq_rec_signal} or $\bm{r}_{n,p}$ in \eqref{eq_rec_signal_after_abf}, depending on whether it is FD or RD. For the RD case, 
\begin{equation} \label{eq_mmse_bf}
    \bm{W}^{(m)} = \left( \bar{\bm{H}}_p \bar{\bm{H}}_p^{H} + N_0 \bm{I}_R \right)^{-1} \bar{\bm{h}}_p^{(m)}
\end{equation}
where $\bar{\bm{H}}_p \triangleq [\bar{\bm{h}}_p^{(1)} \dots \bar{\bm{h}}_p^{(M)}]$. As seen, the channels $\bm{h}_p^{(m)}$ or $\bar{\bm{h}}_p^{(m)}$ are needed for beamformer construction because they are combiners of unknown parameters and should be known. That is why we categorize the MMSE beamformer as an \emph{instantaneous beamformer} since it should be updated with each new instantaneous channel estimate in the rate of channel decorrelation, that is FT-CPI, after each FT training.

There are two modes related to beam tracking, the \emph{Self-Driven} mode and the \emph{Genie-Aided BF} mode. In the proposed system, IEC estimators and beam trackers operate after the beamformers, which are constructed via the previous AoA estimates. Therefore, all the performance measures depend on the previous AoA estimate which describes the \emph{Self-Driven} mode. On the other hand, the \emph{Genie-Aided BF} mode removes the effect of the previous AoA estimate for performance analysis purposes. The beamformers are constructed with true AoAs at the beginning of each ST-CPI, and all the performance measures are collected at the end of each ST-CPI, including angular RMSE from beam trackers although their AoA estimates will not be used in the next ST-CPI. Note that this does not mean that the mismatches due to the movement inside the ST-CPIs are eliminated.

\subsection{Performance Measures} \label{sec_perf_meas}
\subsubsection{Average NMSE for BA-LS} \label{sec_perf_meas_nmse}
The performance of IEC estimation is measured by normalized mean squared error (NMSE), which is defined for the $m$\textsuperscript{th} cluster as\footnote{\label{ft_FTexpect}Note that the expectations both in SINR and NMSE calculations are FT expectations that are conditioned on the settings at the related FT-CPI, such as beamformers and positions.}
\begin{equation} \label{eq_nmse_original}
    \text{NMSE}_p^{(m)} \triangleq \frac{ \mathbb{E} \{ || \hat{\tilde{\bm{h}}}_p^{(m,m)} - \tilde{\bm{h}}_p^{(m,m)}||_2^2 \}}{ \mathbb{E} \{ || \tilde{\bm{h}}_p^{(m,m)} ||_2^2 \}}.
\end{equation}
It is shown in Appendix \ref{app_ba_ls} that the NMSE can be calculated as
\begin{equation} \label{eq_nmse_derived}
    \text{NMSE}_p^{(m)} =  \frac{\text{tr} \left( \tilde{\bm{\Psi}}_p^{(m)} - E^{(m)} \tilde{\bm{R}}_p^{(m,m)} \right)}{E^{(m)} N_{\text{F}} \, \text{tr} \left( \tilde{\bm{R}}_p^{(m,m)} \right)}.
\end{equation}
Then, the \emph{average NMSE} is found by averaging $\text{NMSE}_p^{(m)}$ through all the possible $p$ instants in Monte-Carlo experiments.

\vspace{0.5em}
\subsubsection{Angular RMSE for Beam Tracking}
Angular error is defined as $\hat{\theta}_k^{(\text{ST},m)} - \theta_{(k-1)P+1}^{(m)}$, and the angular root mean squared error (RMSE) is
\begin{equation}
    \text{RMSE}^{(m)} = \sqrt{ \frac{1}{K} \sum_{k=1}^K \left(\hat{\theta}_k^{(\text{ST},m)} - \theta_{(k-1)P+1}^{(m)}\right)^2 }
\end{equation}
for the $m$\textsuperscript{th} cluster, where $K$ is the number of all the possible slow-time instants in Monte-Carlo experiments. Note that the error definition neglects the variation inside the ST-CPI and focuses only on the estimation performance.

\vspace{0.5em}
\subsubsection{Average SINR after ICS-CMF}
The overall performance of the ABF, DBF, beam tracker, and IEC estimator will be measured via SINR at the output of ICS-CMF, which is the symbol estimate $\hat{s}_{n,p}^{(m)}$ given in \eqref{eq_cluster_symbols}. Since the estimated channel is known to the detector rather than the true channel, the true signal term is taken as $S_p^{(m)} \triangleq \sqrt{E^{(m)}} (\hat{\tilde{\bm{h}}}_p^{(m,m)})^{H} \hat{\tilde{\bm{h}}}_p^{(m,m)} s_{n,p}^{(m)}$, and the interference-plus-noise term $N_p^{(m)}=\hat{s}_{n,p}^{(m)} - S^{(m)}$ is the remaining part. With these definitions, SINR for the $m$\textsuperscript{th} cluster is defined as
 \begin{equation}
     \text{SINR}_p^{(m)} \triangleq \mathbb{E}\left\lbrace \frac{|S_p^{(m)}|^2}{|N_p^{(m)}|^2}\right\rbrace.
 \end{equation}
Then, the \emph{average SINR} is found by averaging $\text{SINR}_p^{(m)}$ through all the possible $p$ instants in Monte-Carlo experiments.

\subsection{Simulation Settings} \label{sec_parameters}
Selected parameters for simulations are listed in Table \ref{tab_parameters}. One FT-CPI consists of $N_\text{F}+N_\text{S}=1000$ symbols, which is suitable according to Table \ref{tab_cpis}, and it takes $T_\text{F} =10 \mu s$ assuming a bandwidth of 100 MHz. Selection of $N_\text{F}=10$ and $N_\text{S}=990$ yields 1\% training overhead.

\begin{table}[tb]
\caption{\textsc{Simulation Parameters}}
	\centering
	\resizebox{0.48\textwidth}{!}{%
  \bgroup
\def\arraystretch{1.1}
	\begin{tabular}{|c|c|c|}
		\hline
		Parameter & Description & Value / Details
		\\ \hline \hline
		$N$ & Number of antennas & 128
		\\ \hline
		$M$ & Number of clusters & 4
		\\ \hline
		$U$ & Number of users & 4
		\\ \hline
		$N_{\text{F}}$ & FT training sequence length & 10
		\\ \hline
		$N_{\text{S}}$ & Data mode length in an FT-CPI & 990
		\\ \hline
		$T_{\text{F}}$ & Duration of an FT-CPI & 1e-5 seconds
		\\ \hline
		$P$ & Number of FT-CPIs & $\sim$ 1000
		\\ \hline
		$R$ & Number of RFCs & 16
		\\ \hline
		$D_m$ & Number of DBF outputs & 3
		\\ \hline
		$\theta_0^{(m)}$ & Initial AoAs & \{$10^{\circ}$,$20^{\circ}$,$-10^{\circ}$,$-20^{\circ}$\}
		\\ \hline
		$\Delta^{(m)}$ & ASs & \{$3^{\circ}$,$3^{\circ}$,$3^{\circ}$,$3^{\circ}$\}
		\\ \hline
		$E^{(m)}/N_0$ & Cluster SNRs & \{10,40,30,30\} dB
		\\ 
         \hline
	\end{tabular}
 \egroup
	}
	\label{tab_parameters}			
\end{table}

\begin{figure}[tb]
\centering
\resizebox{0.44\textwidth}{!}{%
	\includegraphics{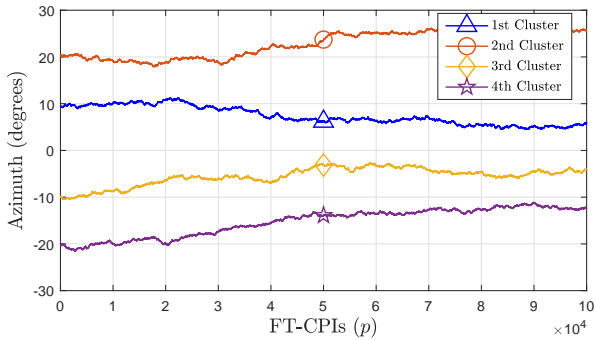}
	}
	\caption{One realization of AoAs from the movement model.}
	\label{fig_movement}
\end{figure}

There are $M=4$ signal clusters. The first cluster, which will be the most interested one, suffers from a significant near-far effect since it is surrounded by 20 to 30 dB stronger clusters. The movement model given in Section \ref{sec_time_var} is implemented with $\sigma_\theta^2=1.45e-4$, $\sigma_\omega^2=1.46e-6$, and initial AoAs given in Table \ref{tab_parameters}. This setting of variances corresponds to 10 m/s speed and 1 m/s\textsuperscript{2} acceleration assuming 150 meters distance from the BS after 1 second of random movement.\footnote{We interpret the standard deviation for the angular position (speed) divided by time as speed (acceleration), in the innovation covariance matrix after 100.000 FT-CPIs (1 second), which is computed similarly to $\bm{\Sigma}_{\nu}^{(\text{ST})}$ in Section  \ref{sec_beamtrack_st_model}.} One example of the resultant movement process is plotted in Fig.~\ref{fig_movement} for a duration of 1 second.\footnote{Note that the selection of $T_\text{F} =10 \mu s$ with the assumption of 100 MHz bandwidth is just to give an insight on a practical operation. $T_\text{F}$ only affects the movement model, and variances would be changed for different $T_\text{F}$ and bandwidth just to have a similar movement with the one in Fig.~\ref{fig_movement}.}

Parameters $D_m$ and $R/M$ should be comparable with the practical rank of effective CCMs $\tilde{\bm{R}}_p^{(m,m)}$, which is determined by the AS $\Delta^{(m)}$ and the number of antennas $N$. If the ratio $R/N$ is limited, the number of active users can be reduced by user scheduling, or $D_m$ and $R/M$ can be reduced by treating the AS $\Delta^{(m)}$ as a design parameter.

For each cluster, only the selected region by ABF $\tilde{\bm{S}}^{(m)}$, which is discussed below \eqref{eq_abf_total}, is searched for BA-ML, OMP, and the periodogram with a resolution of $0.1^{\circ}$, which yields $N_{\theta} \cong 40$. $D_m'=2$ is chosen for SEKF.

\subsection{Numerical Results}
In this section, numerical results from the simulations will be shared. In the simulations, $p_{\text{max}}$ FT-CPIs of movement period, whose one realization for $p_{\text{max}}=1e5$ is given in Fig.~\ref{fig_movement}, is repeated many times to have a Monte-Carlo experiment.\footnote{If any two clusters get closer than $3^{\circ}$, or any cluster gets outside of [$-60^{\circ}$,$60^{\circ}$] region, related realization is ended before $p_{\text{max}}$ FT-CPIs.} In the simulations, the parameters given in Section \ref{sec_parameters} are used, and the performance measures given in Section \ref{sec_perf_meas} are collected. 

\begin{figure}[tb]
\centering
\resizebox{0.44\textwidth}{!}{%
	\includegraphics{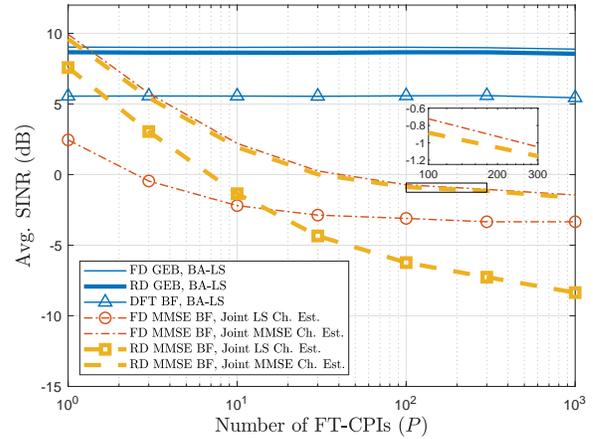}
	}
	\caption{Comparison of instantaneous and statistical beamforming. ($m=1$, $p_{\text{max}}=1000$, Genie-Aided mode)}
	\label{fig_sinr_p}
\end{figure}

\begin{figure}[tb]
\centering
\resizebox{0.44\textwidth}{!}{%
	\includegraphics{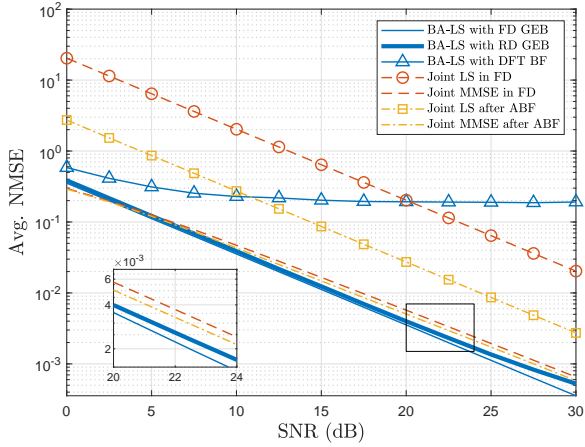}
	}
	\caption{Comparison of instantaneous channel estimation techniques. ($m=1$, $p_{\text{max}}=1000$, Genie-Aided mode)}
	\label{fig_nmse_snr}
\end{figure}

As a beginning, the motivation behind the proposed per-cluster estimation approach with statistical beamforming will be clarified via comparisons with conventional techniques in Figures \ref{fig_sinr_p} and \ref{fig_nmse_snr}. In these figures, mobility and beam tracking is temporarily left out of focus with the settings of Genie-Aided BF mode and $p_{\text{max}}=1000$, for which mobility is almost absent as seen in Fig.~\ref{fig_movement}. In Fig.~\ref{fig_sinr_p}, statistical beamforming (FD-GEB, RD-GEB, and DFT BF) and instantaneous beamforming (FD/RD MMSE BF) are compared according to beamformer update time $P$, which work with second-order statistics (CCMs) and actual channel estimates, respectively. As expected, slow-time alternatives are robust against $P$, while MMSE beamformers fail immediately when $P>1$, since the actual channel decorrelates but statistics remain almost the same. Although the best performance is reached by MMSE beamformers using MMSE channel estimates with $P=1$, the optimum choice in terms of computational complexity per time is statistical beamformers with a large $P$. The first scheme needs MMSE channel estimates for each FT-CPI which is computationally complex as seen from Table \ref{tab_complex}, and the second scheme needs channel estimates from BA-LS, which is simpler. In addition, RD-GEB has less averaged complexity than the RD MMSE beamformer. The simpler alternative of channel estimation for instantaneous beamforming, joint LS, causes a performance loss. Finally, the RD alternatives, which are suitable for HBFs, exhibit minimal performance loss compared with FD alternatives, which are structurally and computationally more complex and needy.

In Fig.~\ref{fig_nmse_snr}, instantaneous channel estimation methods are compared against signal-to-noise ratio (SNR). As discussed in Section \ref{sec_complexity}, BA-LS is computationally simpler than the conventional techniques of joint LS and joint MMSE. Although simpler, it is seen in Fig.~\ref{fig_nmse_snr} that BA-LS does not lose performance compared with the superior technique joint MMSE.\footnote{The vertical axis is \emph{normalized} MSE, and the reason for BA-LS to seem unexpectedly better than joint MMSE is the difference in the sizes of the channels, which are $D_m, D_m, R/M, N, N, R, R$ in the order of the legend.}
This high performance of BA-LS arises from the prior processing via DBF, where GEB is employed. The role of GEB is seen from the performance loss of BA-LS with DFT BF, compared to BA-LS with RD-GEB case. In conclusion, BA-LS with GEB is a very efficient technique in terms of performance and computational complexity. 

\begin{figure}[tb]
\centering
\resizebox{0.44\textwidth}{!}{%
	\includegraphics{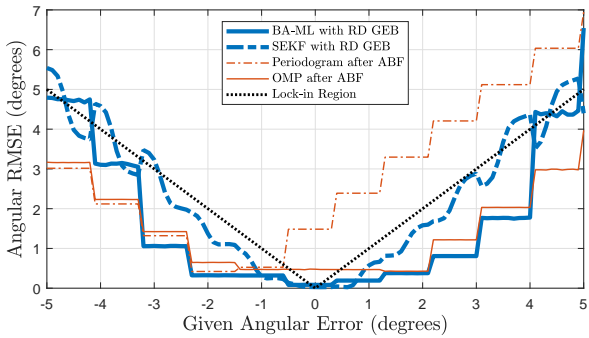}
	}
	\caption{RMS error of angular estimates at the end of an ST-CPI depending on a given error upon which beamformers are constructed at the beginning of the ST-CPI. ($m=1$, $P=1000$)}
	\label{fig_nextAngerr}
\end{figure}

The self-driven mode, which is defined in Section \ref{sec_aux_def} might suffer from instability. Fig.~\ref{fig_nextAngerr} depicts the angular RMSE when the beamformers are constructed with a given error in the AoA. In this regard, Fig.~\ref{fig_nextAngerr} constitutes a basis for the stability of self-driven operation. For a stable self-driven performance, the next angular error is desired to be smaller than the given error, which causes a given error to gradually get smaller and become zero. The region where this behavior is expected is called the \emph{lock-in region}, as shown in the figure. First of all, it is seen that the error in the next estimate is step-like against the previously given error, which is due to discrete angles in the definition of ABF and discretized angular sectors in the definition of RD-GEB. Second, note that the AS is 3 degrees, and $R/M=4$ RFCs of ABF create a beamwidth of approximately 3.5 degrees. Therefore a given error larger than 3 or less than -3 means total beam loss. It is seen in Fig.~\ref{fig_nextAngerr} that BA-ML and OMP offer the most robust performance when there is no beam loss. SEKF also remains in the lock-in region, however, the error due to linearization in \eqref{eq_SEKF_r_model_linear1} and \eqref{eq_SEKF_r_model_linear2} increases with the given angular error. Periodogram fails due to the near-far effect because the cluster $m=2$ is at 20 degrees, which is 30 dB stronger, and estimates deviate towards this cluster. Lastly, we see that the OMP shows nearly 0.5 degrees of RMSE when the given angular error is zero. This is because OMP actually searches for a point target, but BA-ML and SEKF search for a cluster with some AS through their modeling by \eqref{eq_BAML_ch_model} and \eqref{eq_SEKF_r_model}. Therefore, OMP appoints the first powerful AoA candidate inside the AS as an estimate, but others find the center of the observed cluster. 

\begin{figure}[tb]
\centering
\resizebox{0.44\textwidth}{!}{%
	\includegraphics{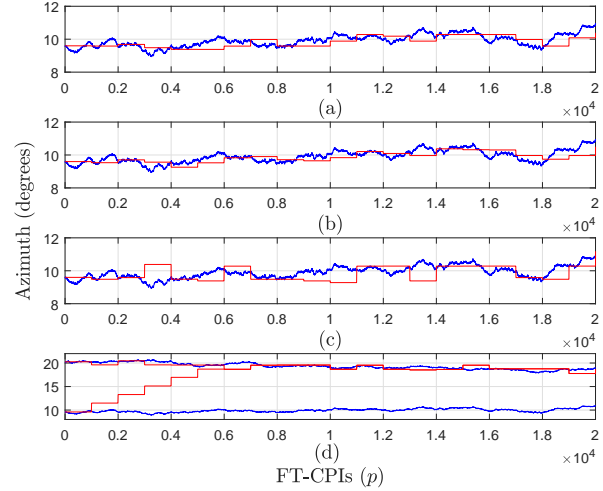}
	}
	\caption{Tracking behaviors of angular estimation techniques: (a) BA-ML with RD-GEB, (b) SEKF with RD-GEB, (c) OMP with ABF, (d) Periodogram with ABF (Blue lines: true AoA, red lines: estimates, $p_{\text{max}}=100000$, $P=1000$, Self-Driven mode).}
	\label{fig_tracking}
\end{figure}

\begin{figure}[tb]
\centering
\resizebox{0.44\textwidth}{!}{%
	\includegraphics{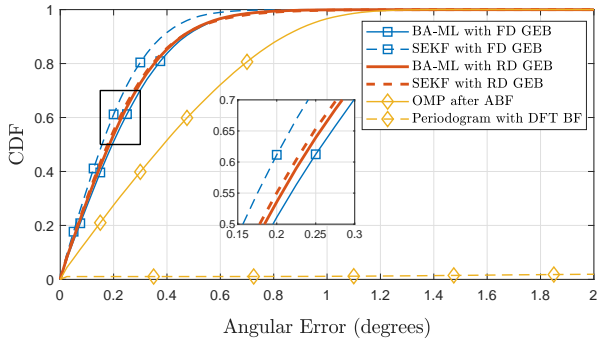}
	}
	\caption{CDF of error in the angular estimates. ($m=1$, $p_{\text{max}}=100000$, $P=1000$, Self-Driven mode)}
	\label{fig_cdf_angerror}     
\end{figure}

\begin{figure}[tb]
\centering
\resizebox{0.44\textwidth}{!}{%
	\includegraphics{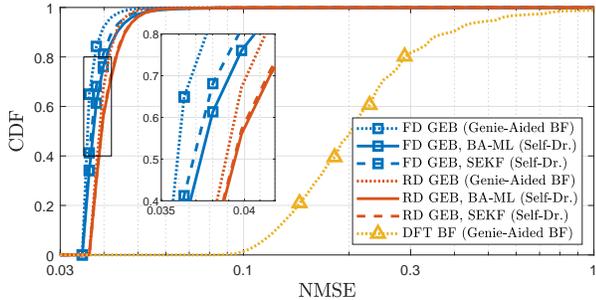}
	}
	\caption{CDF of NMSE for BA-LS with different beamformers. ($m=1$, $p_{\text{max}}=100000$, $P=1000$)}
	\label{fig_cdf_nmse}
\end{figure}

In Figures \ref{fig_tracking}, \ref{fig_cdf_angerror}, and \ref{fig_cdf_nmse}, performances of various angular estimators, the proposed IEC estimator BA-LS, and overall performance measure SINR will be shared. Although some results for the Genie-Aided BF mode are also given for comparison, the main focus is on the performance of the Self-Driven mode with the setting of $p_{\text{max}}=100000$, which leads to more angular variation. In Fig.~\ref{fig_tracking}, tracking behaviors of different angular estimators are given. This figure shows more clearly that the actual AoAs are changing after each FT-CPI, but estimates are produced after every $P$ FT-CPIs, or an ST-CPI. Inside an ST-CPI, beamformers $\bm{S}$ and $\bm{W}^{(m)}$, which are constructed upon these estimates, remain unchanged. It is also seen that the periodogram immediately deviates towards the stronger neighboring cluster, while others are robust to the near-far effect. 

In Fig.~\ref{fig_cdf_angerror}, cumulative distribution functions (CDFs) of angular error for different beam trackers are given. It is seen that the proposed estimators SEKF and BA-ML outperform the alternatives with 90\% of estimates being smaller than 0.5 degrees when operated with FD or RD-GEB. Comparing these with OMP after ABF, we see the contribution of applying cluster-specific DBFs before the estimation, that is the per-cluster approach. With this approach, the complexity is decreased and performance is enhanced. 

In Fig.~\ref{fig_cdf_nmse}, the CDF of NMSE of the proposed instantaneous channel estimation technique BA-LS is plotted. In this figure, the effects of various beamforming and beam tracking techniques on the instantaneous channel estimation accuracy are shown. It is seen that the GEB alternatives used together with the proposed BA-ML and SEKF beam trackers, which are the representatives of the per-cluster estimation approach, result in NMSE values between 0.035 and 0.05. Note that these numbers are lower-bounded by $(\text{SNR} \times N_{\text{F}})^{-1}$, which is 0.01 for $m=1$ and $N_{\text{F}}=10$. It is seen that RD-GEB works with minimal performance loss compared with FD-GEB, and proposed BA-ML and SEKF beam trackers yield NMSE performance similar to Genie-Aided BF mode. Finally, the huge advantage of per-cluster operation is seen from the difference between RD-GEB and DFT BF, which are different only in the design of $\bm{W}^{(m)}$. 

\begin{figure}[tb]
\centering
\resizebox{0.44\textwidth}{!}{%
	\includegraphics{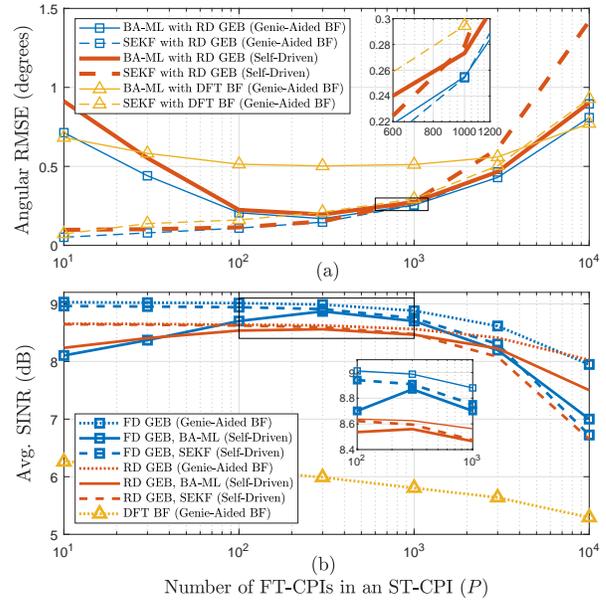}
	}
	\caption{Performances against the length of an ST-CPI: (a) RMS error for angular estimators, (b) Average SINR. ($m=1$, $p_{\text{max}}=100000$)}
	\label{fig_ang_rmse}
\end{figure}

As a final performance measure, Fig.~\ref{fig_ang_rmse} shows angular RMSE and SINR performances against the design parameter $P$, the number of FT-CPIs in an ST-CPI. 
There is a performance loss at large $P$ for all the alternatives due to the loss of angular coherency because of mobility. For small $P$, BA-ML performance degrades due to the decrease in the observation period. However, it is seen that SEKF is not affected by this factor since it is a Kalman filter variant that can collect information from the previous ST-CPIs. Therefore, SEKF can succeed even with the primitive DFT BF, but DFT BF fails in beamforming and channel estimation as seen in subplot (b) and also in Fig.~\ref{fig_cdf_nmse}. Finally, it is seen that the BA-ML and SEKF methods in Self-Driven mode perform similarly to the Genie-Aided mode. Also, the RD alternatives with hybrid beamforming perform with minimal loss compared with the structurally complex and needy FD alternatives.

\section{Conclusion}
In this paper, a novel channel estimation framework is proposed, which includes IEC estimation and beam tracking. This framework, namely the per-cluster estimation with statistical beamforming, is based on the spatial multiplexing of users and signal clusters both in data and training periods. Also, instantaneous channel estimation is repeated in fast-time while beam tracking and beamformer update are slow-time operations. The proposed scheme is shown to be advantageous in terms of both computational complexity and performance compared with the conventional approaches.  

\appendices

\section{Angular Estimate for BA-ML} \label{app_ml}
In this appendix, \eqref{eq_BA_ML} will be derived using \eqref{eq_ba-ml_est_original} and the definitions in \ref{sec_ba-ml}. Since the channel estimates in \eqref{eq_obs_ML} are taken from different FT-CPIs, they are independent and we can use $p(\bm{f}_k^{(m)} | \theta, \{\bm{\beta}_p'\} ) = \prod_{p=1}^{P} p(\hat{\tilde{\bm{h}}}_p^{(m,m)} | \theta, \bm{\beta}_p' )$. Then,
\begin{equation}
    \hat{\theta}_{k+1}^{(\text{ST},m)} = arg \max_{\theta} \prod_{p=1}^{P} \max_{\bm{\beta}_p'} p(\hat{\tilde{\bm{h}}}_p^{(m,m)} | \theta, \bm{\beta}_p' ).
\end{equation}
We assume that $p(\hat{\tilde{\bm{h}}}_p^{(m,m)} | \theta, \bm{\beta}_p' )$ is expressed by $\mathcal{CN} (\bm{E}'(\theta) \bm{\beta}_p',\frac{N_0}{E^{(m)} N_{\text{F}}} \bm{I}_{D_m})$. Then, the likelihood $p(\hat{\tilde{\bm{h}}}_p^{(m,m)} | \theta, \bm{\beta}_p' )$ is maximized for given $\hat{\tilde{\bm{h}}}_p^{(m,m)}$ and $\theta$ by the LS estimate $\hat{\bm{\beta}}_p$ of $\bm{\beta}_p'$. Then, the mean $\bm{E}'(\theta) \hat{\bm{\beta}}_p$ satisfies $\bm{E}'(\theta) \hat{\bm{\beta}}_p = \bm{P}(\theta) \hat{\tilde{\bm{h}}}_p^{(m,m)}$, where 
\begin{equation}
    \bm{P}(\theta) \triangleq \bm{E}'(\theta) \left( \bm{E}'^{H}(\theta) \bm{E}'(\theta) \right)^{-1} \hspace{-3pt} \bm{E}'^{H}(\theta). 
\end{equation}
Then, we can convert the function to log-likelihood to obtain
\begin{align}
    & \begin{aligned}
    \hat{\theta}_{k+1}^{(\text{ST},m)} = arg \min_{\theta} \sum_{p=1}^{P} \left(\hat{\tilde{\bm{h}}}_p^{(m,m)} - \bm{P}(\theta) \hat{\tilde{\bm{h}}}_p^{(m,m)} \right)^{H}
    \\
    \left(\hat{\tilde{\bm{h}}}_p^{(m,m)} - \bm{P}(\theta) \hat{\tilde{\bm{h}}}_p^{(m,m)} \right),
    \end{aligned}
\\ \label{eq_baml_est}
    & \hat{\theta}_{k+1}^{(\text{ST},m)} = arg \min_\theta \sum_{p=1}^{P} (\hat{\tilde{\bm{h}}}_p^{(m,m)})^{H} \bm{M}(\theta) \hat{\tilde{\bm{h}}}_p^{(m,m)},
\end{align}
where $\bm{M}(\theta) \triangleq (\bm{I}_{D_m}-\bm{P}(\theta))^{H} (\bm{I}_{D_m}-\bm{P}(\theta))$, and further $\bm{M}(\theta) = \bm{I}_{D_m} - \bm{P}(\theta)$. The estimator in \eqref{eq_baml_est} can also be implemented as in \eqref{eq_BA_ML} utilizing the properties of the trace operation, which reduces the computational complexity since the second matrix is calculated only once during the search.

\section{Covariance Matrix in SEKF} \label{app_sekf_Q}
The observation $\bm{f}_k^{(m)}$ in \eqref{eq_SEKF_r} can also be written as 
\begin{equation} \label{eq_app_SEKF_f_alternative}
    \bm{f}_k = \frac{1}{P} \sum_{p=(k-1)P+1}^{kP} \hat{\tilde{\bm{h}}}_p^{*} \otimes \hat{\tilde{\bm{h}}}_p,
\end{equation}
where we drop the superscripts for the sake of simplicity. Its mean is $\mathbb{E} \{ \bm{f}_k \} = \text{vec} \{ \bm{R}_f(\theta_k)\}$, and covariance matrix $\bm{Q}_k$ is
\begin{equation} \label{eq_app_SEKF_Q_first}
    \bm{Q}_k = \mathbb{E} \left\{ \bm{f}_k  \bm{f}_k^{H} \right\} -\mathbb{E} \{ \bm{f}_k \} \mathbb{E} \{ \bm{f}_k \}^{H}.
\end{equation}
While the second term is known, the first term should be calculated. Using \eqref{eq_app_SEKF_f_alternative}, it can be written as
\begin{equation} \label{eq_app_SEKF_Eff}
    \mathbb{E} \left\{ \bm{f}_k  \bm{f}_k^{H} \right\} = \frac{1}{P^2} \sum_{p_1} \sum_{p_2} \bm{X}(p_1,p_2)
\end{equation}
where $\bm{X}(p_1,p_2)$ can be written in two equivalent forms:
\begin{align} \label{eq_app_SEKF_X1}
    \bm{X}(p_1,p_2) & = \mathbb{E} \left\{ \left(\hat{\tilde{\bm{h}}}_{p_1}^{*} \otimes \hat{\tilde{\bm{h}}}_{p_1}\right) \left(\hat{\tilde{\bm{h}}}_{p_2}^{*} \otimes \hat{\tilde{\bm{h}}}_{p_2}\right)^{H} \right\}
    \\ \label{eq_app_SEKF_X2}
    \bm{X}(p_1,p_2) & = \mathbb{E} \left\{ \left(\hat{\tilde{\bm{h}}}_{p_1} \hat{\tilde{\bm{h}}}_{p_2}^{H} \right)^{*} \otimes \left(\hat{\tilde{\bm{h}}}_{p_1} \hat{\tilde{\bm{h}}}_{p_2}^{H}\right) \right\}
\end{align}
Note that each entry of $\bm{X}(p_1,p_2)$ is a product of four variables, without summations. Since the channel is complex Gaussian, the identity 
$\mathbb{E} \left\{ a a^{*} b b^{*} \right\} = \mathbb{E} \left\{ a a^{*} \right\} \mathbb{E} \left\{ b b^{*} \right\} + \mathbb{E} \left\{ a b^{*} \right\} \mathbb{E} \left\{ a^{*} b \right\}$ can be used, where $a$ and $b$ are complex Gaussian random variables. The essence of this identity is that different groups are assumed as independent for each term, e.g., $a$ and $b^{*}$ are assumed as independent of $a^{*}$ and $b$ in the second term. The two equivalent forms of $\bm{X}(p_1,p_2)$ in \eqref{eq_app_SEKF_X1} and \eqref{eq_app_SEKF_X2} can be used for these two terms. Consequently,  
\begin{equation} \label{eq_app_SEKF_X_final}
    \bm{X}(p_1,p_2) = \mathbb{E} \{ \bm{f}_k \} \mathbb{E} \{ \bm{f}_k \}^{H} + \delta[p_1-p_2] \bm{R}_f^{*}(\theta_k) \otimes \bm{R}_f(\theta_k)
\end{equation}
Using \eqref{eq_app_SEKF_X_final}, \eqref{eq_app_SEKF_Eff}, and \eqref{eq_app_SEKF_Q_first}, the expression in \eqref{eq_SEKF_Q} is obtained.

\section{Analytical NMSE Expression for BA-LS} \label{app_ba_ls}
For the denominator in the NMSE expression in Section \ref{sec_perf_meas_nmse}, $\mathbb{E}\{ || \tilde{\bm{h}}_p^{(m,m)} ||_2^2 \} = \text{tr} ( \tilde{\bm{R}}_p^{(m,m)} )$ by definition. For the numerator, we define the error vector as $\bm{e} \triangleq \hat{\tilde{\bm{h}}}_p^{(m,m)} - \tilde{\bm{h}}_p^{(m,m)}$, which is evident from \eqref{eq_ch_est_exp}. Its terms are independent and the total energy of $\bm{e}$ is equal to the sum of energies of its terms. We assume that all the symbols have unit power, and $(\bm{s}^{(m)})^{H} \bm{s}^{(m)}$ is deterministic and equal to $N_{\text{F}}$. In addition, $\bm{s}^{(m')}$ are uncorrelated random data for $m'\neq m$, which yields $\mathbb{E} \{ |(\bm{s}^{(m)})^{H} \bm{s}^{(m')}|^2 \}=N_\text{F}$. Then, the energy $\mathbb{E} \{ || \bm{e} ||_2^2 \} = \text{tr} ( \mathbb{E} \{ \bm{e} \bm{e}^{H} \})$ of the error is written as \eqref{app_ba_ls_num1} using \eqref{eq_ch_est_exp}, and simplified as \eqref{app_ba_ls_num2} by inspecting \eqref{eq_Psi_tilde}.
\begin{align} \label{app_ba_ls_num1}
    \mathbb{E} \{ || \bm{e} ||_2^2 \} = &  \sum_{\substack{m'=1 \\ m' \neq m}}^{M} \frac{E^{(m')} N_{\text{F}}}{E^{(m)} N_{\text{F}}^2} \text{tr} \left(\tilde{\bm{R}}_p^{(m,m')} \right) + \text{tr} \left( \frac{N_0 \bm{I}_{D_m}}{E^{(m)} N_{\text{F}}} \right)
\\ \label{app_ba_ls_num2}
    = & \, \frac{1}{E^{(m)} N_{\text{F}}} \, \text{tr} \left( \tilde{\bm{\Psi}}_p^{(m)} - E^{(m)} \tilde{\bm{R}}_p^{(m,m)} \right)
\end{align}


%

    



\ifCLASSOPTIONcaptionsoff
  \newpage
\fi



\bibliography{references} 
\bibliographystyle{IEEEtran}
\end{document}